\documentclass[aip,jmp,amsmath,amssymb,preprint]{revtex4-1}
\usepackage{amssymb,amsmath,bm}

\usepackage{color}
\usepackage{mathrsfs,amsmath}
\usepackage{mathrsfs}
\usepackage{graphicx}
\usepackage{dcolumn}
\usepackage{bm}
\usepackage{natbib}
\usepackage{amsthm}
\usepackage{ulem}

\begin{document}
\newtheorem{thm}{Theorem}[section]
\newtheorem{lemma}[thm]{Lemma}
\newtheorem{prop}[thm]{Proposition}
\newtheorem{rem}[thm]{Remark}
\newtheorem{cor}[thm]{Corollary}

\title{Large Deviations Principle for the Fluctuating Boltzmann Equation}
\date{\today} 

\author{Liu Hong}
\email{hongliu@sysu.edu.cn}
\affiliation{School of Mathematics, Sun Yat-sen University, Guangzhou, 510275, P.R.C.}          

\begin{abstract}
The Boltzmann equation is one of the most famous equations and has vast applications  in modern science. In the current study, we take the randomness of binary collisions into consideration and generalize the classical Boltzmann equation into a stochastic framework. The corresponding Kolmogorov forward equations and Liouville equation in either discrete or continuous time and state space are derived respectively, whose characteristic line gives the Boltzmann equation as a consequence of the law of large numbers. Then the large deviations principle for these equations is established, which not only explains the probabilistic origin of the H-theorem in the Boltzmann equation, but also provides a natural way to incorporate the Boltzmann equation into a broader Hamiltonian structure. The so-called Hamilton-Boltzmann equation enjoys many significant merits, like time reversibility, the conservation laws of mass, momentum and energy, Maxwellian-Boltzmann distribution as the equilibrium solution, etc. We also present results under the diffusive limit in parallel. Finally, the macroscopic hydrodynamic models including 13 moments are derived with respect to our Hamilton-Boltzmann equation under the BGK approximation. We expect our study can inspire new insights into the classical Boltzmann equation from either the stochastic aspect or a Hamiltonian view.
\end{abstract}


\maketitle

\section{Introduction}

The Boltzmann equation, as it is named, was first derived by Ludwig E. Boltzmann is 1872 \cite{Boltzmann-1872} and later has been applied by the same author to study the diffusion coefficient, heat conduction rate \textit{etc.} during a transportation process in 1875 \cite{Boltzmann-1875}. Due to its elegant mathematical properties, like conservation laws of density, momentum and energy, the H-theorem, as well as its close relation to the classical hydrodynamic equations, like the Euler equations and Navier-Stocks equations through the Chapman-Enskog expansion, the Boltzmann equation plays a key role in the kinetic theory of non-equilibrium thermodynamics, and also has been extensively applied to the study of gas dynamics, heat conduction, neutron scattering, plasma kinetics, \textit{etc.} \cite{Cercignani-1989}

A general form of the classical Boltzmann equation reads\cite{Cercignani-1989}
\begin{eqnarray}
	\frac{\partial f}{\partial t}+\vec{v}\cdot\frac{\partial f}{\partial\vec{r}}+\vec{F}(\vec{r})\cdot\frac{\partial f}{\partial\vec{v}}=\int\int d\vec{v}_1d\vec{\omega}B(\vec{\omega},\vec{v},\vec{v}_1) \big(f'f_1'-f f_1\big),
\end{eqnarray}
where $f=f(\vec{r},\vec{v},t)$ represents the particle number density at position $(\vec{r},\vec{v})$ in the so-called $\mu$-space, which is $2d$-dimensional and includes both position and velocity as its coordinates. In the same way, we note $f_1=f(\vec{r},\vec{v}_1,t)$, $f'=f(\vec{r},\vec{v}',t)$ and $f_1'=f(\vec{r},\vec{v}_1',t)$ for short. $\vec{F}(\vec{r})$ stands for the external force, which is assumed to be a function of $\vec{r}$. The integral on the right-hand side is the famous binary-collision term, which is originated from random collisions between two particles sharing the same location in space. Consequently, $\vec{v}$ and $\vec{v}_1$ represent velocities of two-particles before collision, while $\vec{v}'$ and $\vec{v}_1'$ are those after collision. Due to the conservation of momentum and kinetic energy during each collision, they satisfy $\vec{v}+\vec{v}_1=\vec{v}'+\vec{v}_1'$ and $\vec{v}^2+\vec{v}_1^2=\vec{v}'^2+\vec{v}_1'^2$. $B(\vec{\omega},\vec{v},\vec{v}_1)=B(\vec{\omega},\vec{v}_1,\vec{v})=B(\vec{\omega},\vec{v}',\vec{v}_1')\geq0$ is the collision kernel, in which $\vec{w}=(\theta,\phi)$ denotes the scattering angle.

The global well-posedness of solutions to the Boltzmann equation is a central issue in the field of modern partial differential equations. For general initial values, R.J. Diperna and P.L. Lions proved the global existence of renormalized solutions to the Boltzmann equation for the first time\cite{DiPerna-1989}. However, the solution's uniqueness and regularity has been an open problem till now. Another significant progress was made by L. Desvillettes and C. Villani in 2005\cite{Desvillettes-2005}, who showed the solution to the Boltzmann equation will converge to its equilibrium state in the long time, by assuming the solution is uniformly bounded in a Sobolev space with sufficiently high regularity.

In this study, we are going to construct an explicit stochastic process, which includes the classical Boltzmann equation as its deterministic limit. In other words, the Boltzmann equation emerges a result of laws of large numbers, while of course our stochastic framework contains far more fruitful results. For example, referring to our framework, the results for Gaussian fluctuations around the solution of the Boltzmann equation studied by Spohn \cite{Spohn-1991}, Gaspard \cite{Gaspard-2013} and \textit{et al.} can be reproduced by considering the central limit theorem; while the rigorous mathematical derivations by Rezakhanlou \cite{Rezakhanlou-2004} and Bodineau et al. \cite{Bodineau-2017, Bodineau-2020} on dilute gas dynamics for hard spheres and in the Boltzmann-Grad limit are understood as consequences of large deviation asymptotics in a more unified way. Our framework is closely related to the Kac's model \cite{Kac-1956, Mischler-2013}, which is a continuous time Markov chain mimicking the homogeneous
Boltzmann's equation. In contrast, our standing point is a discrete time Markov chain on discrete $\mu$-space lattice, which includes both position and velocity and thus is heterogenous.

With the explicit stochastic model in hand, we have examined its large deviation rate function, which satisfies an integral-differential Hamilton-Jacobi equation, by referring to the famous Freidlin-Wentzel's theorem for small random perturbations of a dynamical system \cite{Freidlin-1998}. The stationary solution of the Hamilton-Jacobi equation turns to be the classical Boltzmann entropy. In this way, the H-theorem can be understood as a natural mathematical consequence of the large deviation principle rather than a coincidence.

Furthermore, time-reversible Hamiltonian dynamics, which contains the Boltzmann equation as a special realization with zero momentum, is constructed with respect to the Hamilton-Jacobi equation. This elegant construction has at least two significant merits. On the one hand, it provides an insightful understanding on the origin of time-irreversible Boltzmann dynamics from the Hamiltonian dynamics, which is a long unsolved issue and has bothered mathematician and physicist for a hundred of years; on the other hand, the newly-derived Hamilton equations may offer a wider and better mathematical framework, like equipped with a Hamiltonian structure, Lagrangian variational principle and integrable theory, for studying dilute gas dynamics and fluid mechanics, which previously are regarded as dissipative and time irreversible from the view of Boltzmann's formulation. Maybe the global well-posedness of solutions to the Boltzmann equation can be considered from this aspect too. Considering the potential significance of the extended Hamiltonian dynamics, we have further studied its basic mathematical properties from several aspects: the equilibrium solution, conservation laws and entropy balance equation.

The whole paper is organized as follows: In section II, with respect to a lattice view of $\mu$ space, we reformulate the binary collision of particles into a Markov chain. The corresponding Kolmogorov forward equations in both discrete time and state space, master equations in continuous time but discrete state space, and the Liouville equation in the continuous limit are constructed sequentially. Especially, the classical Boltzmann equation appears as the characteristic line of the Liouville equation. In Section III, we apply the large deviations principle to the master equations, from which the Hamilton-Jacobi equation for the large deviations rate function is derived. And the stationary large deviations rate function, which takes the famous Boltzmann entropy as a special solution, turns to the relative entropy function for the Boltzmann equation. Furthermore, with respect to the HJE, its corresponding Hamiltonian dynamics is established, whose  entropy balance equation, conservation laws and equilibrium solution are discussed separately. Section IV is a natural generalization of above results to the diffusive limit. The last section contains a conclusion and some brief discussions.

\section{Boltzmann's Formulation in a Stochastic Framework}
To avoid potential difficulties on the rigorous justification of PDEs, we firstly turn to a discrete-lattice-based view and then take the continuum limit to reach our final goal. Rather than dealing with the whole $\mu$-space (position plus velocity), here what we actually considered is a bounded closed subspace by assuming the particles are constrained within a finite region and with finite velocity. Without loss of generality, we further suppose it is a rectangle with $L_s$ as the maximal length and $L_v$ as the maximal magnitude of velocity. Therefore, the system size is given by $V=L_s^d\times L_v^d$, where $d$ stands for the dimension.

Then we equally divide the whole system into $M_s^d\times M_v^d$ cells. And each cell will be denoted by $[\vec{r}_i,\vec{v}_k]$, where $i\in[1,M_s^d],k\in[1,M_v^d]$ are indexes for partitions on space and velocity coordinates respectively. Without loss of generality, we let $\vec{r}_i$ and $\vec{v}_k$ be the central position and velocity in each cell. In accordance with the local equilibrium assumption in classical irreversible thermodynamics \cite{Groot-1962}, each cell should be microscopically large enough to allow making meaningful statistical averaging and macroscopically adequately small so that within each one the variations in particle position and velocity are negligible. Mathematically, it means $a\ll L_s/M_s\ll L$ and $L_v/M_v\ll L_v$, where $a$ is the radius of a particle.

Referring to above setup, now we denote the total particle number in the system as $N$, while the vector of occupancy number (or the particle number in each cell) as $\vec{n}(t)=(n[\vec{r}_i,\vec{v}_k](t))$. Then the changes in the occupancy number will come from two different sources according to the Boltzmann's picture: a direct transportation between neighboring cells and binary collisions which happen between two particles sharing the same position. The former is a common constitutive part for general hydrodynamic equations based on conservation laws, while the latter is a unique piece to the Boltzmann equation and is also the origin of the H-theorem. In contrast to the original deterministic formulation of Boltzmann, here we are going to reconsider these two sources in a stochastic framework.

\subsection{Deterministic Transportation}
Since in the Boltzmann's picture the $\mu$-space includes both position and velocity, there are two kinds of direct transportation, which are both deterministic (or with probability as one in a stochastic description). One is the position shift due to non-zero velocity, the other is the velocity shift due to external forces.

For the former, the net change in occupancy number per each transportation process from cell $[\vec{r}_i,\vec{v}_k]$ to cell $[\vec{r}_j,\vec{v}_l]$ during time interval $dt$ at time $t$ is given by
\begin{eqnarray}
d\vec{n}_v(\vec{r}_i,\vec{v}_k,\vec{r}_j,\vec{v}_l;t,dt)=n[\vec{r}_i,\vec{v}_k](t)\Gamma_2(\vec{r}_i,\vec{v}_k;\vec{r}_j,\vec{v}_l)\delta(\vec{v}_k-\vec{v}_l)\delta(\vec{r}_j-\vec{r}_i-\vec{v}_kdt),
\end{eqnarray}
where $\Gamma_2(\vec{r}_i,\vec{v}_k;\vec{r}_j,\vec{v}_l)$ is a vector with index running from $1$ to $M_s^d\times M_v^d$. Except for $\Gamma_2[\vec{r}_i,\vec{v}_k]=-1$ and $\Gamma_2[\vec{r}_j,\vec{v}_l]=1$, all of its remaining components are zeros. In particular, we define  $\Gamma_2(\vec{r}_i,\vec{v}_k;\vec{r}_i,\vec{v}_k)=0$,  which leads to  $d\vec{n}_v(\vec{r}_i,\vec{v}_k,\vec{r}_i,\vec{v}_k;t,dt)=0$ in consistence with the conservation law of particle number. $\delta(\vec{v}_k-\vec{v}_l)$ is the Dirac's delta function, which equals to one if and only if $\vec{v}_k=\vec{v}_l$. Similarly, the net change in occupancy number from cell $[\vec{r}_i,\vec{v}_k]$ to cell $[\vec{r}_j,\vec{v}_l]$ during time interval $dt$ caused by external forces is
\begin{eqnarray}
d\vec{n}_f(\vec{r}_i,\vec{v}_k,\vec{r}_j,\vec{v}_l;t,dt)=n[\vec{r}_i,\vec{v}_k](t)\Gamma_2(\vec{r}_i,\vec{v}_k;\vec{r}_j,\vec{v}_l)\delta(\vec{r}_j-\vec{r}_i)\delta(\vec{v}_l-\vec{v}_k-\vec{F}(\vec{r}_i)dt).
\end{eqnarray}

As a consequence, the net change in occupancy number for each cell by all possible direct transportation  during time interval $dt$ can be calculated as
\begin{eqnarray}
&&d\vec{n}_{tra}(t,dt)=\sum_{\vec{r}_i,\vec{v}_k,\vec{r}_j,\vec{v}_l}\big[d\vec{n}_v(\vec{r}_i,\vec{v}_k,\vec{r}_j,\vec{v}_l;t,dt)+d\vec{n}_f(\vec{r}_i,\vec{v}_k,\vec{r}_j,\vec{v}_l;t,dt)\big]\nonumber\\
&=&\sum_{\vec{r}_i,\vec{v}_k}\bigg\{n[\vec{r}_i,\vec{v}_k](t)\Gamma_2(\vec{r}_i,\vec{v}_k;\vec{r}_i+\vec{v}_kdt,\vec{v}_k)+
n[\vec{r}_i,\vec{v}_k](t)\Gamma_2(\vec{r}_i,\vec{v}_k;\vec{r}_i,\vec{v}_k+\vec{F}(\vec{r}_i)dt)\bigg\}.
\end{eqnarray}
In particular, if we focus on the rate of occupancy number change in cell $[\vec{r}_i,\vec{v}_k]$ at time $t$ by direct transportation, it gives
\begin{eqnarray}
&&\lim_{\substack{dt\rightarrow0\\ M_s,M_v\rightarrow\infty\\ M_s,M_v\sim O(dt^{-1}) }}\frac{dn_{tra}[\vec{r}_i,\vec{v}_k](t,dt)}{CV\cdot dt}\nonumber\\
&=&\lim_{\substack{dt\rightarrow0\\ M_s,M_v\rightarrow\infty\\ M_s,M_v\sim O(dt^{-1})}}\bigg\{\frac{n[\vec{r}_i-\vec{v}_kdt,\vec{v}_k](t)-n[\vec{r}_i,\vec{v}_k](t)}{CV\cdot dt}+\frac{n[\vec{r}_i,\vec{v}_k-\vec{F}(\vec{r}_i)dt](t)-n[\vec{r}_i,\vec{v}_k](t)}{CV\cdot dt}\bigg\}\nonumber\\
&=&-\vec{v}\cdot\frac{\partial f(\vec{r},\vec{v},t)}{\partial\vec{r}}-\vec{F}(\vec{r})\cdot\frac{\partial f(\vec{r},\vec{v},t)}{\partial\vec{v}},
\end{eqnarray}
in the continuum limit. $CV=(L_s/M_s)^d\times(L_v/M_v)^d$ denotes the cell volume. And the particle density function reads $f(\vec{r},\vec{v},t)=\lim_{M_s,M_v\rightarrow\infty}n[\vec{r}_i,\vec{v}_k](t)/CV$, where $\vec{r}=\lim_{M_s\rightarrow\infty}\vec{r}_i$ and $\vec{v}=\lim_{M_v\rightarrow\infty}\vec{v}_k$. The exact meaning of above continuum limit is that we require the partition of the time and space becomes finer and finer, meanwhile their relative ratio has to be kept constant. So that we have $L_s/M_s\sim O(dt)$, $L_v/M_v\sim O(dt)$.

\subsection{Stochastic Formulation of Binary Collisions}
Binary collision happens between two particles sharing the same position, so that only their velocities are changed during the collision, \textit{i.e.} $\vec{v}_k,\vec{v}_l\rightarrow\vec{v}_{k'},\vec{v}_{l'}$. Note the velocities before and after collision must obey conservation laws of momentum and kinetic energy, which means $\vec{v}_k+\vec{v}_l=\vec{v}_{k'}+\vec{v}_{l'}$ and $\vec{v}_k^2+\vec{v}_l^2=\vec{v}_{k'}^2+\vec{v}_{l'}^2$. In contrast to Boltzmann's original deterministic formulation, here we consider the particle binary collisions happen in a random way. And the cellular forward and reverse rates for particle binary collisions in cell $[\vec{r}_i,\vec{v}_k]$ and cell $[\vec{r}_j,\vec{v}_l]$ are given by
\begin{eqnarray}
&&R_c^+[\vec{n}](\vec{r}_i,\vec{v}_k,\vec{r}_j,\vec{v}_l,\vec{\omega}_m;t)=(4\epsilon)^{-1}B(\vec{\omega}_m,\vec{v}_k,\vec{v}_l) n[\vec{r}_i,\vec{v}_k](t)n[\vec{r}_j,\vec{v}_l](t)\delta(\vec{r}_i-\vec{r}_j),\\
&&R_c^-[\vec{n}](\vec{r}_i,\vec{v}_k,\vec{r}_j,\vec{v}_l,\vec{\omega}_m;t)=(4\epsilon)^{-1}B(\vec{\omega}_m,\vec{v}_k,\vec{v}_l) n[\vec{r}_{i},\vec{v}_{k'}](t)n[\vec{r}_{j},\vec{v}_{l'}](t)\delta(\vec{r}_i-\vec{r}_j),
\end{eqnarray}
where $\vec{\omega}_m$ denotes the discrete scattering angle between the incoming and outcoming particles. $\epsilon=CV/V=(M_sM_v)^{-d}\ll1$ is a dimensionless small parameter.

Correspondingly, vectors for the net change in occupancy number due to forward or reverse particle binary collision in cell $[\vec{r}_i,\vec{v}_k]$ and cell $[\vec{r}_j,\vec{v}_l]$ are
\begin{eqnarray} &&d\vec{n}_c^+(\vec{r}_i,\vec{v}_k,\vec{r}_j,\vec{v}_l,\vec{\omega}_m)=-d\vec{n}_c^-(\vec{r}_i,\vec{v}_k,\vec{r}_j,\vec{v}_l,\vec{\omega}_m)=\epsilon\cdot\Gamma_4(\vec{r}_i,\vec{v}_k,\vec{r}_j,\vec{v}_l,\vec{\omega}_m),
\end{eqnarray}
where $\Gamma_4(\vec{r}_i,\vec{v}_k,\vec{r}_j,\vec{v}_l,\vec{\omega}_m\neq\vec{0})$ is a vector of length $M^d\times N^d$, of which only four components are nonzero, \textit{i.e.} $\Gamma_4[\vec{r}_i,\vec{v}_k]=\Gamma_4[\vec{r}_j,\vec{v}_l]=-1$ and $\Gamma_4[\vec{r}_i,\vec{v}_{k'}]=\Gamma_4[\vec{r}_j,\vec{v}_{l'}]=1$. Again, we require $\Gamma_4(\vec{r}_i,\vec{v}_k,\vec{r}_j,\vec{v}_l,\vec{0})=0$, meaning during the collision these two particles just simply exchange their velocities and thus can not cause any change in the occupancy number.

The inclusion of particle's binary collision is closely related to the dilute-gas limit, which states the particle density should be kept so low that the many-body particle collision (like three-body or four-body) has a rare chance to happen. Mathematically, a gas is said dilute if we have the following relations for scale separation hold $a\ll\rho^{-1/d}\ll l_f$, where $a$ is the particle radius (or cross section to be exact), $\rho$ is the average particle density, $l_f$ is the mean free path of particles or particle's average travelling
distance before a collision.

\begin{figure}
	\centering
	\includegraphics[width=0.8\textwidth, angle=0]{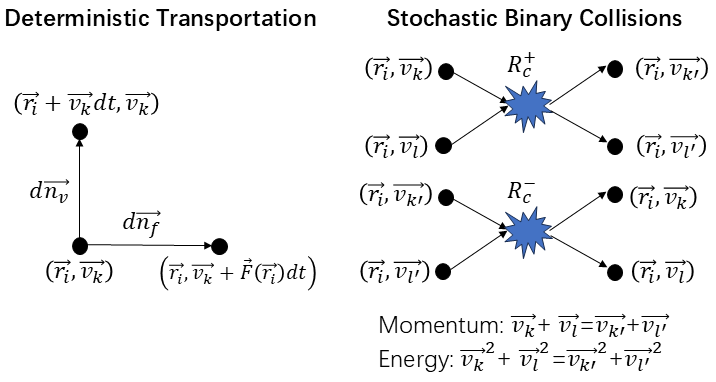}	
	\caption{Illustration on the particle transportation and binary collisions in the stochastic formulation of Boltzmann equation.}
	\label{fig_1}%
\end{figure}

\subsection{Fluctuating Boltzmann Equation and Its Characteristic Line}
With respect to two kinds of elementary processes -- direction transportation and binary collision discussed above, the transition probability of occupancy number changing from $\vec{n}'$ to $\vec{n}$ per time interval $dt$ can be calculated according to the following ansatz
\begin{eqnarray}
&&Q(\vec{n}',t|\vec{n}=\vec{n}'+d\vec{n}_{tra}(t,dt)+d\vec{n},t+dt)\nonumber\\
&=&\left\{
\begin{array}{l}
R_c^+[\vec{n}']dt+o(dt),\; if\;d\vec{n}=d\vec{n}_c^+(\vec{r}_i,\vec{v}_k,\vec{r}_j,\vec{v}_l,\vec{\omega}_m\neq\vec{0}),\\
R_c^-[\vec{n}']dt+o(dt),\; if\;d\vec{n}=d\vec{n}_c^-(\vec{r}_i,\vec{v}_k,\vec{r}_j,\vec{v}_l,\vec{\omega}_m\neq\vec{0}),\\
1-\sum_{\vec{r}_i,\vec{v}_k,\vec{r}_j,\vec{v}_l,\vec{\omega}_m\neq\vec{0}}(R_c^{+}[\vec{n}']+R_c^{-}[\vec{n}'])dt+o(dt),\;if\;d\vec{n}=0,\\
0,\;for\; other\; d\vec{n}.
\end{array}
\right.
\end{eqnarray}

Having the transition probability in hand, the probability $P(\vec{n},t)$ satisfies the Kolmogorov forward equation,
\begin{eqnarray}
P(\vec{n},t+dt)&=&\sum_{\vec{n}'}Q(\vec{n}',t|\vec{n},t+dt)P(\vec{n}',t)=P(\vec{n}-d\vec{n}_{tra}(t,dt),t)\nonumber\\
&+&\sum_{\vec{r}_i,\vec{v}_k,\vec{r}_j,\vec{v}_l,\vec{\omega}_m\neq\vec{0}}\big\{R_c^{+}[\vec{n}-d\vec{n}_{tra}(t,dt)-d\vec{n}_c^+]P(\vec{n}-d\vec{n}_{tra}(t,dt)-d\vec{n}_c^+,t)\nonumber\\
&&\qquad\qquad\qquad-R_c^{+}[\vec{n}-d\vec{n}_{tra}(t,dt)]P(\vec{n}-d\vec{n}_{tra}(t,dt),t)\big\}dt\nonumber\\
&+&\sum_{\vec{r}_i,\vec{v}_k,\vec{r}_j,\vec{v}_l,\vec{\omega}_m\neq\vec{0}}\big\{R_c^{-}[\vec{n}-d\vec{n}_{tra}(t,dt)-d\vec{n}_c^-]P(\vec{n}-d\vec{n}_{tra}(t,dt)-d\vec{n}_c^-,t)\nonumber\\
&&\qquad\qquad\qquad-R_c^{-}[\vec{n}-d\vec{n}_{tra}(t,dt)]P(\vec{n}-d\vec{n}_{tra}(t,dt),t)\big\}dt.
\end{eqnarray}


In the continuum limit, by taking Taylor expansion of $P(\vec{n},t)$ up to the first order, the Kolmogorov forward equation turns to the following Liouville equation
\begin{eqnarray*}
\frac{\partial P(f,t)}{\partial t}+\int\int d\vec{r}d\vec{v}\bigg[-\vec{v}\cdot\frac{\partial f}{\partial\vec{r}}-\vec{F}(\vec{r})\cdot\frac{\partial f}{\partial\vec{v}}+\int\int B(\vec{\omega},\vec{v},\vec{v}_1) \big(f'f_1'-f f_1\big)d\vec{v}_1d\vec{\omega}\bigg]\frac{\delta P(f,t)}{\delta f}=0,
\end{eqnarray*}
where $P(f,t)=\lim_{M_s,M_v\rightarrow\infty}P(\vec{n},t)$ is a functional of $f(\vec{r},\vec{v},t)$, $B(\vec{\omega},\vec{v},\vec{v}_1)=\lim_{M_s,M_v\rightarrow\infty}B(\vec{\omega}_m,\vec{v}_k,\vec{v}_l)/(\vec{\omega}_{m+1}-\vec{\omega}_m)$.

It is worthy noting that the characteristic line of the above Liouville equation exactly gives the Boltzmann equation \cite{Evans-2010},
\begin{eqnarray*}
	\frac{\partial f}{\partial t}+\vec{v}\cdot\frac{\partial f}{\partial\vec{r}}+\vec{F}(\vec{r})\cdot\frac{\partial f}{\partial\vec{v}}=\int\int d\vec{v}_1d\vec{\omega}B(\vec{\omega},\vec{v},\vec{v}_1) \big(f'f_1'-f f_1\big),
\end{eqnarray*}
which clearly is a consequence of laws of large numbers. Therefore, the above Kolmogorov forward equation and Liouville equation both can be regarded as stochastic generalizations of the classical Boltzmann equation. And we call them as the fluctuating Boltzmann equation in either discrete or continuous form.

\section{Large Deviation Principle and Extended Hamiltonian Dynamics for the Boltzmann Equation}

\subsection{WKB Method and Hamilton-Jacobi Equation for the Large Deviation Rate Function}
Analogous to the study of Ge and Qian on the Poisson process of chemical reactions in the large volume limit \cite{Ge-2017}, here we look into the large deviation asymptotics of the fluctuating Boltzmann equation. By inserting a WKB ansatz $P(\vec{n},t)\simeq \exp[-I(\vec{n},t)/\epsilon]$ into the Kolmogorov forward equation, where $I(\vec{n},t)$ is the large deviation rate function and $\epsilon=CV/V\ll1$ is a small parameter, and applying the Taylor expansion, it is straightforward to see that
\begin{eqnarray*}
&&\frac{\exp[-I(\vec{n},t+dt)/\epsilon]-\exp[-I(\vec{n}-d\vec{n}_{tra}(t,dt),t)/\epsilon]}{dt}\\
&=&\exp(-I/\epsilon)\frac{\exp\big[-(\partial I/\partial t)dt/\epsilon-(\partial I/\partial \vec{n})d\vec{n}_{tra}(t,dt)/\epsilon+o(dt)\big]-1}{dt}\bigg|_{I=I(\vec{n}-d\vec{n}_{tra}(t,dt),t)}\\
&=&-\frac{\exp(-I/\epsilon)}{\epsilon}\bigg[\frac{\partial I}{\partial t}+\frac{\partial I}{\partial \vec{n}}\frac{d\vec{n}_{tra}(t,dt)}{dt}+o(1)\bigg]\bigg|_{I=I(\vec{n}-d\vec{n}_{tra}(t,dt),t)}\\
&=&\exp(-I/\epsilon)\sum_{\vec{r}_i,\vec{v}_k,\vec{r}_j,\vec{v}_l,\vec{\omega}_m\neq\vec{0}}R_c^{+}\bigg\{\bigg[1+d\vec{n}_c^+\bigg(\frac{\partial R_c^{+}}{\partial \vec{n}}\bigg)+o(\epsilon)\bigg]\exp\bigg[\frac{d\vec{n}_c^+(\partial I/\partial \vec{n})+o(\epsilon)}{\epsilon}\bigg]-1\bigg\}\\
&&+\exp(-I/\epsilon)\sum_{\vec{r}_i,\vec{v}_k,\vec{r}_j,\vec{v}_l,\vec{\omega}_m\neq\vec{0}}R_c^{-}\bigg\{\bigg[1+d\vec{n}_c^-\bigg(\frac{\partial R_c^{-}}{\partial \vec{n}}\bigg)+o(\epsilon)\bigg]\exp\bigg[\frac{d\vec{n}_c^-(\partial I/\partial \vec{n})+o(\epsilon)}{\epsilon}\bigg]-1\bigg\},
\end{eqnarray*}
where $I=I(\vec{n}-d\vec{n}_{tra}(t,dt),t), R_c^{\pm}=R_c^{\pm}[\vec{n}-d\vec{n}_{tra}(t,dt)]$. Based on it, we derive the evolution equation for the large deviation rate function,
\begin{eqnarray*}
&&\frac{\partial I(\vec{n},t)}{\partial t}+\frac{\partial I(\vec{n},t)}{\partial \vec{n}}\frac{d\vec{n}_{tra}(t,dt)}{dt}\\
&=&-\epsilon\sum_{\vec{r}_i,\vec{v}_k,\vec{r}_j,\vec{v}_l,\vec{\omega}_m\neq\vec{0}}R_c^{+}[\vec{n}]\bigg\{\exp\bigg[\frac{1}{\epsilon}d\vec{n}_c^+\frac{\partial I(\vec{n},t)}{\partial \vec{n}}+o(1)\bigg]-1\bigg\}\\
&&-\epsilon\sum_{\vec{r}_i,\vec{v}_k,\vec{r}_j,\vec{v}_l,\vec{\omega}_m\neq\vec{0}}R_c^{-}[\vec{n}]\bigg\{\exp\bigg[\frac{1}{\epsilon}d\vec{n}_c^-\frac{\partial I(\vec{n},t)}{\partial \vec{n}}+o(1)\bigg]-1\bigg\}+o(1)\\
&=&-\sum_{\vec{r}_i,\vec{v}_k,\vec{r}_j,\vec{v}_l,\vec{\omega}_m\neq\vec{0}}\tilde{R}_c^{+}[\vec{n}]\bigg\{\exp\bigg[\Gamma_4\frac{\partial I(\vec{n},t)}{\partial \vec{n}}+o(1)\bigg]-1\bigg\}\\
&&-\sum_{\vec{r}_i,\vec{v}_k,\vec{r}_j,\vec{v}_l,\vec{\omega}_m\neq\vec{0}}\tilde{R}_c^{-}[\vec{n}]\bigg\{\exp\bigg[-\Gamma_4\frac{\partial I(\vec{n},t)}{\partial \vec{n}}+o(1)\bigg]-1\bigg\}+o(1),
\end{eqnarray*}
where $\tilde{R}_c^{\pm}[\vec{n}]=\epsilon R_c^{\pm}[\vec{n}]$.

In the continuum limit $M_s,M_v\rightarrow\infty, dt\rightarrow0$ and $M_s,M_v\sim O(dt^{-1})$, it means the large deviation rate function as a functional of $f$  satisfies the following Hamilton-Jacobi equation (HJE)
\begin{eqnarray}
&&\frac{\partial I}{\partial t}+\int\int d\vec{r}d\vec{v}\bigg[-\vec{v}\cdot\frac{\partial f}{\partial\vec{r}}-\vec{F}(\vec{r})\cdot\frac{\partial f}{\partial\vec{v}}\bigg]\frac{\delta I}{\delta f}\nonumber\\
&=&-\frac{1}{4}\int\int\int\int d\vec{r}d\vec{v} d\vec{v}_1d\vec{\omega}B(\vec{\omega},\vec{v},\vec{v}_1) \bigg\{f'f_1'\bigg[\exp\bigg(\frac{\delta I}{\delta f}+\frac{\delta I}{\delta f_1}-\frac{\delta I}{\delta f'}-\frac{\delta I}{\delta f_1'}\bigg)-1\bigg]\nonumber\\
&&+ff_1\bigg[\exp\bigg(-\frac{\delta I}{\delta f}-\frac{\delta I}{\delta f_1}+\frac{\delta I}{\delta f'}+\frac{\delta I}{\delta f_1'}\bigg)-1\bigg]\bigg\},
\end{eqnarray}
by noticing  $I=I(f,t)=\lim_{M_s,M_v\rightarrow\infty}I(\vec{n},t)$. Due to the invariance of the collision kernel $B(\vec{\omega},\vec{v},\vec{v}_1)=B(\vec{\omega},\vec{v}',\vec{v}_1')$, the term on the right-hand side could be shortened as $-\frac{1}{2}\int\int\int\int d\vec{r}d\vec{v} d\vec{v}_1d\vec{\omega}B(\vec{\omega},\vec{v},\vec{v}_1) ff_1\big[\exp\big(-\delta I/\delta f-\delta I/\delta f_1+\delta I/\delta f'+\delta I/\delta f_1'\big)-1\big]$.

A key step in above derivation is the renormalization of particle's binary collision rates and occupancy number change by $\epsilon$, which allows the elimination of $\epsilon$ in the zeroth-order HJE as well as a proper limit during the continuation procedure.

\subsection{Boltzmann Entropy as a Stationary Solution of HJE}

It is well-known that the stationary solution of HJE $I^{ss}[f]=I(f,\infty)$ serves as the Lyapunov function for the deterministic dynamics or the characteristic line of the Liouville equation, since
\begin{eqnarray*}
&&\frac{dI^{ss}[f(\vec{r},\vec{v},t)]}{dt}=\int\int d\vec{r}d\vec{v}\frac{\delta I^{ss}}{\delta f}\frac{\partial f}{\partial t}\nonumber\\
&=&\int\int d\vec{r}d\vec{v}\frac{\delta I^{ss}}{\delta f}\bigg[-\vec{v}\cdot\frac{\partial f}{\partial\vec{r}}-\vec{F}(\vec{r})\cdot\frac{\partial f}{\partial\vec{v}}+\int\int d\vec{v}_1d\vec{\omega}B(\vec{\omega},\vec{v},\vec{v}_1) \big(f'f_1'-f f_1\big)\bigg]\\
&=&-\frac{1}{4}\int\int\int\int d\vec{r}d\vec{v}d\vec{v}_1d\vec{\omega}B(\vec{\omega},\vec{v},\vec{v}_1) \bigg\{f'f_1'\bigg[\exp\bigg(\frac{\delta I^{ss}}{\delta f}+\frac{\delta I^{ss}}{\delta f_1}-\frac{\delta I^{ss}}{\delta f'}-\frac{\delta I^{ss}}{\delta f_1'}\bigg)-1\bigg]\nonumber\\
&&+ff_1\bigg[\exp\bigg(-\frac{\delta I^{ss}}{\delta f}-\frac{\delta I^{ss}}{\delta f_1}+\frac{\delta I^{ss}}{\delta f'}+\frac{\delta I^{ss}}{\delta f_1'}\bigg)-1\bigg]-\big(f'f_1'-f f_1\big)\bigg(\frac{\delta I^{ss}}{\delta f}+\frac{\delta I^{ss}}{\delta f_1}-\frac{\delta I^{ss}}{\delta f'}-\frac{\delta I^{ss}}{\delta f_1'}\bigg)\bigg\}\nonumber\\
&=&-\frac{1}{4}\int\int\int\int d\vec{r}d\vec{v}d\vec{v}_1d\vec{\omega}B(\vec{\omega},\vec{v},\vec{v}_1) \big\{f'f_1'\big[\exp(\psi)-1-\psi\big]+ff_1\big[\exp(-\psi)-1+\psi\big]\big\}\nonumber\\
&\leq&0,
\end{eqnarray*}
in which $\psi=\delta I^{ss}/\delta f+\delta I^{ss}/\delta f_1-\delta I^{ss}/\delta f'-\delta I^{ss}/\delta f_1'$. The last step is due to the fact $e^\psi\geq1+\psi$.

In the absence of external forces $\vec{F}=0$, the stationary solution to the Hamilton-Jacobi equation can be explicitly solved,
\begin{eqnarray*}
\bigg[\frac{f'}{y'}\frac{f_1'}{y_1'}-\frac{f}{y}\frac{f_1}{y_1}\bigg](y'y_1'-yy_1)=0,
\end{eqnarray*}
which gives either $y=const$ or $f/y=const$ with $y=\exp(\delta I^{ss}/\delta f)$. The former leads to $I^{ss}[f]=c_0\int f(\vec{r},\vec{v},t)d\vec{v}$, while the latter gives the famous Boltzmann entropy\footnote{Strictly speaking, $\int\int d\vec{r}d\vec{v} f(\vec{r},\vec{v},t)\ln f(\vec{r},\vec{v},t)$ is a stationary solution to the HJE, but not necessarily the large deviation rate function, since it may not fulfill the positiveness requirement. It is straightforward to check that another stationary solution of HJE, which is also the large deviation rate function, is given by $\int\int d\vec{r}d\vec{v} f(\vec{r},\vec{v},t)\ln [f(\vec{r},\vec{v},t)/f^{eq}(\vec{r},\vec{v},t)]$, where $f^{eq}(\vec{r},\vec{v},t)$ stands for the Maxwell-Boltzmann distribution.},
\begin{eqnarray}
\label{Boltzmann-entropy}
I^{ss}[f(\vec{r},\vec{v},t)]=\int\int d\vec{r}d\vec{v} f(\vec{r},\vec{v},t)\ln f(\vec{r},\vec{v},t).
\end{eqnarray}
This result clarifies the origin of the Boltzmann entropy function and the H theorem, which emerge as a consequence of the large deviation principle of a much broader stochastic model -- the fluctuation Boltzmann equation to be exact.

%

\subsection{Hamilton-Boltzmann Equations}
With respect to the HJE, we can introduce a Hamiltonian function
\begin{eqnarray}
\label{Hamilton}
H(f,y)&=&-\int\int d\vec{r}d\vec{v}\bigg[\vec{v}\cdot\frac{\partial f}{\partial\vec{r}}+\vec{F}(\vec{r})\cdot\frac{\partial f}{\partial\vec{v}}\bigg]y \nonumber\\
&&+\frac{1}{2}\int\int\int\int d\vec{r}d\vec{v} d\vec{v}_1d\vec{\omega}B(\vec{\omega},\vec{v},\vec{v}_1)ff_1\big[\exp\big(-y-y_1+y'+y_1'\big)-1\big],
\end{eqnarray}
in which the generalized momentum is defined as $y=\delta I(f,t)/\delta f$. Clearly, the first term in the Hamiltonian function is originated from transportation, while the second term is associated with binary collision of particles.

Now with the Hamiltonian function in hand, it becomes possible to extend the classical irreversible Boltzmann equation to time-reversible Hamiltonian dynamics in infinite dimension,
\begin{eqnarray*}
\frac{\partial f}{\partial t}=\frac{\delta H}{\delta y}&=&-\vec{v}\cdot\frac{\partial f}{\partial\vec{r}}-\vec{F}(\vec{r})\cdot\frac{\partial f}{\partial\vec{v}}+\int\int d\vec{v}_1d\vec{\omega}B(\vec{\omega},\vec{v},\vec{v}_1) \big[f'f_1'\exp\big(y+y_1-y'-y_1'\big)\nonumber\\
&&-ff_1\exp\big(-y-y_1+y'+y_1'\big)\big],\\
\frac{\partial y}{\partial t}=-\frac{\delta H}{\delta f}&=&-\vec{v}\cdot\frac{\partial y}{\partial\vec{r}}-\vec{F}(\vec{r})\cdot\frac{\partial y}{\partial\vec{v}}-\int\int d\vec{v}_1d\vec{\omega}B(\vec{\omega},\vec{v},\vec{v}_1)f_1\big[\exp\big(-y-y_1+y'+y_1'\big)-1\big].
\end{eqnarray*}
By introducing a new material differentiation operator $D/Dt=\partial/\partial t+\vec{v}\cdot\partial /\partial\vec{r}+\vec{F}(\vec{r})\cdot\partial/\partial\vec{v}$, above equations could be rewritten into a more compact way,
\begin{eqnarray}
&&\frac{Df}{Dt}=\int\int d\vec{v}_1d\vec{\omega}B(\vec{\omega},\vec{v},\vec{v}_1) \big[f'f_1'\exp\big(y+y_1-y'-y_1'\big)-ff_1\exp\big(-y-y_1+y'+y_1'\big)\big],\\
&&\frac{Dy}{Dt}=-\int\int d\vec{v}_1d\vec{\omega}B(\vec{\omega},\vec{v},\vec{v}_1)f_1\big[\exp\big(-y-y_1+y'+y_1'\big)-1\big],
\end{eqnarray}
which is a group of coupled integral-differential equations. And we call it the Hamilton-Boltzmann equations (HBEs). The HBEs can be rewritten into a more compact and anti-symmetric form, that is
\begin{eqnarray}
	&&\frac{Dg}{Dt}=\int\int d\vec{v}_1d\vec{\omega}B(\vec{\omega},\vec{v},\vec{v}_1) h_1(g'g_1'-gg_1),\\
	&&\frac{Dh}{Dt}=-\int\int d\vec{v}_1d\vec{\omega}B(\vec{\omega},\vec{v},\vec{v}_1)g_1(h'h_1'-hh_1),
\end{eqnarray}
where $h=\exp(y),g=f/h=\exp(-y)f$. Notice this is a canonical transformation, since 
\begin{equation}
y\frac{\partial f}{\partial t}-H(f,y)=h\frac{\partial g}{\partial t}-H(g,h)+\frac{\partial}{\partial t}[(\ln(h)-1)gh].
\end{equation}

So by incorporating the fluctuations $y$ in a momentum space, we precisely compensate the free energy dissipation in the original Boltzmann equation and make it becoming non-dissipative and fully time-reversible. And it is easy to see that the classical Boltzmann equation turns to be a special case of the above Hamiltonian dynamics by setting $y=0$. This result also shows that the path characterized by the Boltzmann equation corresponds to the most probable one among all possible paths given by the HBEs, which again is a consequence of laws of large numbers.

\subsection{Entropy Balance Equation}
The H-theorem plays a central role in characterizing the time-irreversible nature of the classical Boltzmann equation, while in the HBEs we will see that it will be replaced by a more general entropy balance equation. With respect to the entropy function $S(t)=\int\int (f-f\ln f)d\vec{r}d\vec{v}$.
\begin{eqnarray}
\frac{dS(t)}{dt}&=&\int\int \frac{D}{Dt}(f-f\ln f)d\vec{r}d\vec{v}=-\int\int \frac{D}{Dt}f\cdot\ln fd\vec{r}d\vec{v}\nonumber\\
&=&\frac{1}{4}\int\int\int\int d\vec{r}d\vec{v}d\vec{v}_1d\vec{\omega}B(\vec{\omega},\vec{v},\vec{v}_1) \big\{f'f_1'\big[\exp\big(y+y_1-y'-y_1'\big)-1\big]\nonumber\\
&&-ff_1\big[\exp\big(-y-y_1+y'+y_1'\big)-1\big]\big\}\ln\bigg(\frac{f'f_1'}{ff_1}\bigg)\nonumber\\
&&+\frac{1}{4}\int\int\int\int d\vec{r}d\vec{v}d\vec{v}_1d\vec{\omega}B(\vec{\omega},\vec{v},\vec{v}_1) (f'f_1'-ff_1)\ln\bigg(\frac{f'f_1'}{ff_1}\bigg),
\end{eqnarray}
in which the first term is recognized as the entropy flux, while the second term is the entropy production rate, which is always non-negative.

\subsection{Conservation Laws}
The HBEs enjoy many remarkable merits as the classical Boltzmann equation, one of which is the conserved quantities expressed through conservation laws. In the absence of external forces ($\vec{F}(\vec{r})=0$), it is straightforward to see that the three summational invariants during binary collision \cite{Grad-1949} -- the total particle number $N(t)=\int\int f d\vec{r}d\vec{v}$, momentum $P_0(t)=\int\int f\vec{v} d\vec{r}d\vec{v}$ and kinetic energy $E(t)=\int\int (f\vec{v}^2/2) d\vec{r}d\vec{v}$, where $d$ stands for the space dimension, are still conserved in the HBEs. Their corresponding conservation laws read
\begin{eqnarray}
\frac{dN(t)}{dt}&=&\int\int \frac{D}{Dt}\big(f\big)d\vec{r}d\vec{v}=0,\\
\frac{dP_0(t)}{dt}&=&\int\int \frac{D}{Dt}\big(f\vec{v}\big)d\vec{r}d\vec{v}=0,\\
\frac{dE(t)}{dt}&=&\int\int \frac{D}{Dt}\bigg(\frac{f\vec{v}\cdot\vec{v}}{2}\bigg)d\vec{r}d\vec{v}=0,
\end{eqnarray}
in which $D/Dt=\partial/\partial t+\vec{v}\cdot\partial /\partial\vec{r}$. Besides them, since the Hamiltonian function does not explicitly depend on time, it is time-conserved too ($dH(f,y)/dt=0$).

\subsection{Specific Solutions}
In the absence of external forces ($\vec{F}(\vec{r})=0$), a specific solution of HBEs reads
\begin{eqnarray}
	f^{eq}(\vec{r},\vec{v},t)&=&\frac{\rho}{(2\pi \theta)^{d/2}}\exp\bigg[-\frac{(\vec{v}-\vec{u})^2}{2\theta}\bigg],\\
	y^{eq}(\vec{r},\vec{v},t)&=&a_0(t,\vec{r})+\vec{a}_1(t,\vec{r})\cdot\vec{v}+a_2(t,\vec{r})\vec{v}\cdot\vec{v},
\end{eqnarray}
where $a_0, \vec{a}_1, a_2$ are parameters to be determined through initial conditions. $\rho(\vec{r},t)=\int f d\vec{v}$, $\rho\vec{u}(\vec{r},t)=\int f\vec{v} d\vec{v}$ and $d\rho \theta(\vec{r},t)/2=\int f(\vec{v}-\vec{u})^2/2 d\vec{v}$ represent the local particle number density, momentum and kinetic energy separately. It is noted that in this specific solution $f^{eq}(\vec{r},\vec{v},t)$ is completely consistent with the Maxwell-Boltzmann distribution, while $y^{eq}(\vec{r},\vec{v},t)$ is a combination of three collision invariants.

\section{The Diffusive Limit}

\subsection{Jump Moments and Diffusion Approximation}
In this part, we are going to investigate the diffusion approximation of the fluctuation Boltzmann equation. With respect to the transition probability, the first jump moment is
\begin{eqnarray}
&&M_1(\vec{n},t;dt)=\sum_{\vec{n}'}(\vec{n}'-\vec{n})Q(\vec{n},t|\vec{n}',t+dt)\nonumber\\
&=&d\vec{n}_{tra}(t,dt)+\sum_{\vec{r}_i,\vec{v}_k,\vec{r}_j,\vec{v}_l,\vec{\omega}_m}(R_c^{+}[\vec{n}]dn_c^++R_c^{-}[\vec{n}]dn_c^-)(\vec{r}_i,\vec{v}_k,\vec{r}_j,\vec{v}_l,\vec{\omega}_m;t)dt.
\end{eqnarray}
In the last term, the constraint $\vec{\omega}_m\neq\vec{0}$ is removed since $\Gamma_4(\vec{r}_i,\vec{v}_k,\vec{r}_j,\vec{v}_l,\vec{0})=0$ as we required.

Similarly, the second jump moment can be calculated as
\begin{eqnarray}
&&M_2(\vec{n},t;dt)=\sum_{\vec{n}'}(\vec{n}'-\vec{n})(\vec{n}'-\vec{n})^TQ(\vec{n},t|\vec{n}',t+dt)\nonumber\\
&=&d\vec{n}_{tra}(t,dt) d\vec{n}_{tra}(t,dt)^T+\sum_{\vec{r}_i,\vec{v}_k,\vec{r}_j,\vec{v}_l,\vec{\omega}_m}(R_c^{+}[\vec{n}]dn_c^++R_c^{-}[\vec{n}]dn_c^-)(\vec{r}_i,\vec{v}_k,\vec{r}_j,\vec{v}_l,\vec{\omega}_m;t)dt d\vec{n}_{tra}(t,dt)^T\nonumber\\
&&+\sum_{\vec{r}_i,\vec{v}_k,\vec{r}_j,\vec{v}_l,\vec{\omega}_m}d\vec{n}_{tra}(t,dt)(R_c^{+}[\vec{n}]dn_c^++R_c^{-}[\vec{n}]dn_c^-)^T(\vec{r}_i,\vec{v}_k,\vec{r}_j,\vec{v}_l,\vec{\omega}_m;t)dt\nonumber\\
&&+\sum_{\vec{r}_i,\vec{v}_k,\vec{r}_j,\vec{v}_l,\vec{\omega}_m}[dn_c^+(R_c^{+}[\vec{n}]+R_c^{-}[\vec{n}])(dn_c^+)^T](\vec{r}_i,\vec{v}_k,\vec{r}_j,\vec{v}_l,\vec{\omega}_m;t)dt,\nonumber\\
&=&\sum_{\vec{r}_i,\vec{v}_k,\vec{r}_j,\vec{v}_l,\vec{\omega}_m}[dn_c^+(R_c^{+}[\vec{n}]+R_c^{-}[\vec{n}])(dn_c^+)^T](\vec{r}_i,\vec{v}_k,\vec{r}_j,\vec{v}_l,\vec{\omega}_m;t)dt+o(dt).
\end{eqnarray}
The last step holds as $d\vec{n}_{tra}(t,dt)\sim dt$.

In the same way, we can go to high-order jump moments, which in general are nonzero too. However, by a simple scaling examination, it is easily seen that high-order jump moments will be negligibly small when comparing to first and second jump moments as $\epsilon\rightarrow0$. In fact, according to our formulation, the first jump moment will be of $O(1)$, the second jump moment is of $O(\epsilon)$. Moments of third and above orders will be of at least $O(\epsilon^{2})$.

\subsection{The Diffusive-Fluctuating Boltzmann Equation}
With the first and second jump moments in hand, it is straightforward to show that the probability $P(\vec{n},t)$ satisfies
\begin{eqnarray*}
\frac{P(\vec{n},t+dt)-P(\vec{n},t)}{dt}=-\frac{\partial}{\partial \vec{n}}\cdot\bigg\{\frac{M_1(\vec{n},t;dt)}{dt}P(\vec{n},t)-\frac{\partial}{\partial \vec{n}}\cdot\bigg[\frac{M_2(\vec{n},t;dt)}{dt}P(\vec{n},t)\bigg]\bigg\}+o(\epsilon),
\end{eqnarray*}
by Kramers-Moyal expansion. Further taking the continuum limit, we will arrive at the Fokker-Planck equation
\begin{eqnarray}
	\label{FP-eq}
	\frac{\partial p(f,t)}{\partial t}=-\int\int d\vec{r}d\vec{v}\frac{\delta}{\delta f}\bigg\{u(f)p(f,t)-\int d\vec{v_2}\frac{\delta}{\delta f_2}\big[\epsilon D(f,f_2)p(f,t)\big]\bigg\},
\end{eqnarray}
where the drift and diffusion coefficients are given by
\begin{eqnarray}
u(f)&=&\lim_{\substack{dt\rightarrow0\\ M_s,M_v\rightarrow\infty\\ M_s,M_v\sim O(dt^{-1})}}\frac{M_1(\vec{n},t;dt)}{CV\cdot dt}\nonumber\\
&=&-\vec{v}\cdot\frac{\partial f}{\partial\vec{r}}-\vec{F}(\vec{r})\cdot\frac{\partial f}{\partial\vec{v}}+\int\int d\vec{v}_1d\vec{\omega}B(\vec{\omega},\vec{v},\vec{v}_1) \big(f'f_1'-ff_1\big),\\
D(f,f_2)&=&\lim_{\substack{dt\rightarrow0\\ M_s,M_v\rightarrow\infty\\ M_s,M_v\sim O(dt^{-1})}}\frac{M_2(\vec{n},t;dt)}{\epsilon CV^2\cdot dt}=\frac{1}{2}\int\int d\vec{v_1}d{\vec{\omega}}B(\vec{\omega},\vec{v},\vec{v}_1) \big(f'f_1'+ff_1\big)\nonumber\\
&&\times\big[\delta(\vec{v}-\vec{v}_2)+\delta(\vec{v}_1-\vec{v}_2)-\delta(\vec{v}'-\vec{v}_2)-\delta(\vec{v}_1'-\vec{v}_2)\big],
\end{eqnarray}
with $f_2=f(\vec{r},\vec{v_2},t)$.

The diffusion coefficient $D(f,f_2)$ is both symmetric and semi-positive definite. The former property is easily seen from its expression, while the latter property holds as
\begin{eqnarray*}
&&\int\int\int d\vec{r}d\vec{v}d\vec{v_2} g(f)D(f,f_2)g(f_2)\\
&=&\frac{1}{2}\int\int\int\int d\vec{r}d\vec{v}d\vec{v_1}d{\vec{\omega}}B(\vec{\omega},\vec{v},\vec{v}_1) \big(f'f_1'+ff_1\big)g(f)\big[g(f)+g(f_1)-g(f')-g(f_1')\big]\\
&=&\frac{1}{8}\int\int\int\int d\vec{r}d\vec{v}d\vec{v_1}d{\vec{\omega}}B(\vec{\omega},\vec{v},\vec{v}_1) \big(f'f_1'+ff_1\big)\big[g(f)+g(f_1)-g(f')-g(f_1')\big]^2\geq0,
\end{eqnarray*}
for any two arbitrary functionals $g(f)$ and $g(f_2)$.

It is well-known that there exists a Langevin equation closely related to the above Fokker-Planck equation in the sense of It$\hat{o}$'s integral (though not unique in general), which reads
\begin{eqnarray}
	df=u(f)dt+\sqrt{2\epsilon D(f,f_2)}\cdot dW_t,
\end{eqnarray}
where $W_t$ is an infinite-dimensional Wiener process. According to Joel Keizer \cite{Keizer-1987}, we call it the diffusive-fluctuation Boltzmann equation, since it does give the classical Boltzmann equation by neglecting the last fluctuation term. It is worthy to notice that the diffusive-fluctuating Boltzmann equation is actually a weak-noise perturbation of the deterministic Boltzmann equation, since the fluctuation term scales as $O(\sqrt{\epsilon})$.

\subsection{Large Deviation Rate Function and Hamiltonian Dynamics}
According to Freidlin and Wentzell \cite{Freidlin-1998}, there exists a large deviation rate function
\begin{eqnarray}
\tilde{I}(f,t)=-\lim_{\epsilon\rightarrow0}\epsilon\ln p(f,t)
\end{eqnarray}
associated with above non-dimensionalized diffusive-fluctuating Boltzmann equation. Referring to the WKB method, it is straightforward to see that the LDRF satisfies the following Hamilton-Jacobi equation,
\begin{eqnarray}
\frac{\partial \tilde{I}(f,t)}{\partial  t}=-\int\int\int d\vec{r}d\vec{v}d\vec{v_2}\bigg[\frac{\delta \tilde{I}(f,t)}{\delta f}\bigg] D(f,f_2)\bigg[\frac{\delta \tilde{I}(f,t)}{\delta f_2}\bigg]-\int\int d\vec{r}d\vec{v}\bigg[\frac{\delta \tilde{I}(f,t)}{\delta f}\bigg] u(f).
\end{eqnarray}
It is easy to check that the stationary solution of above HJE $\tilde{I}^{ss}(f)=\tilde{I}(f,\infty)$ serves as the Lyapunov function for the deterministic dynamics or the classical Boltzmann equation too, since
\begin{eqnarray*}
\frac{d\tilde{I}^{ss}(f)}{dt}&=&\int\int d\vec{r}d\vec{v}\bigg[\frac{\delta \tilde{I}^{ss}(f)}{\delta f}\bigg]\frac{\partial f}{\partial t}
=\int\int d\vec{r}d\vec{v}\bigg[\frac{\delta \tilde{I}^{ss}(f)}{\delta f}\bigg] u(f)\\
&=&-\int\int\int d\vec{r}d\vec{v}d\vec{v_2}\bigg[\frac{\delta \tilde{I}^{ss}(f)}{\delta f}\bigg] D(f,f_2)\bigg[\frac{\delta \tilde{I}^{ss}(f)}{\delta f_2}\bigg]\leq0.
\end{eqnarray*}
The last inequality holds as $D(f)$ is both symmetric and semi-positive definite. Interestingly, under the diffusion approximation, the classical Boltzmann entropy function is no longer a stationary solution to the HJE.

With respect to the HJE, we introduce the Hamiltonian function as
\begin{eqnarray}
\tilde{H}(f,y)&=&\int\int d\vec{r}d\vec{v}[u(f)y]+\int\int\int d\vec{r}d\vec{v}d\vec{v_2}[yD(f,f_2)y_2]\nonumber\\
&=&\int\int d\vec{r}d\vec{v}\bigg[-\vec{v}\cdot\frac{\partial f}{\partial\vec{r}}-\vec{F}(\vec{r})\cdot\frac{\partial f}{\partial\vec{v}}+\int\int  d\vec{v}_1d\vec{\omega}B(\vec{\omega},\vec{v},\vec{v}_1)(f'f_1'-ff_1)\bigg]y \nonumber\\
&&+\frac{1}{8}\int\int\int\int d\vec{r}d\vec{v} d\vec{v}_1d\vec{\omega}B(\vec{\omega},\vec{v},\vec{v}_1)(f'f_1'+ff_1)\big(y+y_1-y'-y_1'\big)^2,
\end{eqnarray}
in which the generalized momentum is defined as $y=\delta \tilde{I}(f,t)/\delta f$. Therefore, the first term in the Hamiltonian function is identified as the kinetic energy, while the second term is the potential energy. Note, by performing Taylor expansion of $\exp(y+y_1-y'-y_1')$ up to the second order when $y+y_1-y'-y_1'\approx0$, this Hamiltonian function can be derived from the full one in Eq. (\ref{Hamilton}) .

Now with the Hamiltonian function in hand, it becomes possible to extend the classical irreversible Boltzmann equation to time-reversible Hamiltonian dynamics under the diffusion approximation,
\begin{eqnarray}
&&\frac{df}{dt}=\frac{\delta \tilde{H}}{\delta y}=u(f)+\int d\vec{v_2}D(f,f_2)y_2,\\
&&\frac{dy}{dt}=-\frac{\delta \tilde{H}}{\delta f}=-\frac{\delta}{\delta f}\int\int\int d\vec{r}d\vec{v}d\vec{v_2}[yD(f,f_2)y_2]-\frac{\delta}{\delta f}\int\int d\vec{r}d\vec{v}[u(f)y].
\end{eqnarray}
By inserting the explicit formulas for $u(f)$ and $D(f,f_2)$, we obtain
\begin{eqnarray}
	\frac{Df}{Dt}&=&\int\int d\vec{v}_1d\vec{\omega}B(\vec{\omega},\vec{v},\vec{v}_1) \bigg[\big(f'f_1'-ff_1\big)+\big(f'f_1'+ff_1\big)\big(y+y_1-y'-y_1'\big)\bigg],\\
	\frac{Dy}{Dt}&=&\int\int d\vec{v}_1d\vec{\omega}B(\vec{\omega},\vec{v},\vec{v}_1) f_1(y+y_1-y'-y_1')\bigg[1-\frac{1}{2}(y+y_1-y'-y_1')\bigg].
\end{eqnarray}
In above, we write $f=f(\vec{r},\vec{v},t)$, $f_1=f(\vec{r},\vec{v}_1,t)$, $f'=f(\vec{r},\vec{v}',t)$, $f_1'=f(\vec{r},\vec{v}_1',t)$, and $y=y(\vec{r},\vec{v},t)$, $y_1=y(\vec{r},\vec{v}_1,t)$, $y'=y(\vec{r},\vec{v}',t)$, $y_1'=y(\vec{r},\vec{v}_1',t)$ for short. So by incorporating the fluctuations $y$ in a momentum space, we precisely compensate the free energy dissipation in the original Boltzmann equation and make it becoming non-dissipative and fully time-reversible.


\section{The Hydrodynamic Limit and Macroscopic Equations}
\subsection{Derivation of the classical BGK model}
The BGK model provides an important simplification of the classical Boltzmann equation, especially for its complicated collision integral on the right-hand side \cite{Bhatnagar-1954}. Although this model does not contain any physical relevant feature about binary collisions, it still has raised the wide interest of the academic community, since it contains most of the basic properties
of hydrodynamics: conservation laws of mass, momentum and energy; the H theorem about entropy; and a proper fluid limit leading to the compressible Euler equations, etc. \cite{Kremer-2010}

To get the BGK model, we can start with the following special Boltzmann equation,
\begin{eqnarray}
\label{Boltzmann-1}
\frac{\partial f}{\partial t}+\vec{v}\cdot\frac{\partial f}{\partial \vec{r}}=\int\int d\vec{v_1}{\vec{\omega}}B(\vec{\omega},\vec{v},\vec{v_1})[(f')^{eq}(f'_1)^{eq}-ff_1],
\end{eqnarray}
by putting equilibration requirement on the distribution function of particles after collision. $(f')^{eq}=f^{eq}(\vec{r},\vec{v}',t)=\rho^f(2\pi \theta^f)^{-3/2}\exp[-(\vec{v}'-\vec{u}^f)^2/(2\theta^f)]$ stands for the Maxwell-Boltzmann distribution for the equilibrium state  in three dimensional space, where  $\rho^f(\vec{r},t)=\int f d\vec{v}$, $\rho^f\vec{u}^f(\vec{r},t)=\int f\vec{v} d\vec{v}$ and $3\rho^f \theta^f(\vec{r},t)/2=\int f(\vec{v}-\vec{u}^f)^2/2 d\vec{v}$ represent the local particle number density, momentum and kinetic energy separately. Inserting the concrete forms of $(f')^{eq}$ and $(f'_1)^{eq}$ into Eq. (\ref{Boltzmann-1}), we have
\begin{eqnarray*}
&&\frac{\partial f}{\partial t}+\vec{v}\cdot\frac{\partial f}{\partial \vec{r}}\\
&=&\int\int d\vec{v_1}d{\vec{\omega}}B(\vec{\omega},\vec{v},\vec{v_1})\bigg\{\frac{(\rho^f)^2}{(2\pi \theta^f)^3}\exp\bigg[\frac{-(\vec{v}'-\vec{u}^f)^2-(\vec{v}_1'-\vec{u}^f)^2}{2\theta^f}\bigg]-ff_1\bigg\}\\
&=&\int\int d\vec{v_1}d{\vec{\omega}}B(\vec{\omega},\vec{v},\vec{v_1})\bigg\{\frac{(\rho^f)^2}{(2\pi \theta^f)^3}\exp\bigg[\frac{-(\vec{v}-\vec{u}^f)^2-(\vec{v}_1-\vec{u}^f)^2}{2\theta^f}\bigg]-ff_1\bigg\}\\
&=&f^{eq}\int\int d\vec{v_1}d{\vec{\omega}}B(\vec{\omega},\vec{v},\vec{v_1})f_1^{eq}-f\int\int d\vec{v_1}d{\vec{\omega}}B(\vec{\omega},\vec{v},\vec{v_1})f_1\\
&=&f^{eq}\int d\vec{v_1}f_1^{eq}\cdot\int d{\vec{\omega}}B(\vec{\omega})-f\int d\vec{v_1}f_1\cdot\int d{\vec{\omega}}B(\vec{\omega})=\frac{\rho^f}{\sigma}(f^{eq}-f),
\end{eqnarray*}
where the collision frequency $\sigma^{-1}=\int d{\vec{\omega}}B(\vec{\omega})$ is a constant by assuming the collision kernel only depends on the scattering angle $B(\vec{\omega},\vec{v},\vec{v_1})=B(\vec{\omega})$. In the second step during above derivation, we use the momentum and energy conservation laws during particle binary collision, i.e. $\vec{v}+\vec{v}_1=\vec{v}'+\vec{v}'_1$ and $\vec{v}^2+\vec{v}_1^2=(\vec{v}')^2+(\vec{v}'_1)^2$. The last equation is known as the BGK model.

\subsection{BGK approximation for the Hamilton-Boltzmann equations}
Similar to the Boltzmann equation, we adopt the BGK approximation for the Hamilton-Boltzmann equations by replacing the first part in the integral with their equilibrium values,
\begin{eqnarray}
	&&\frac{Dg}{Dt}=\int\int d\vec{v}_1d\vec{\omega}B(\vec{\omega}) [h_1^{eq}(g')^{eq}(g_1')^{eq}-h_1gg_1],\\
	&&\frac{Dh}{Dt}=-\int\int d\vec{v}_1d\vec{\omega}B(\vec{\omega})[g_1^{eq}(h')^{eq}(h_1')^{eq}-g_1hh_1],
\end{eqnarray}
where $D/Dt=\partial/\partial t+\vec{v}\cdot \partial/\partial \vec{r}$. The stationary solutions are indicated by the supscript 	``$^{eq}$''. For example, we have $(g')^{eq}=g^{eq}(\vec{r},\vec{v}',t)=\rho^g(2\pi \theta^g)^{-3/2}\exp[-(\vec{v}'-\vec{u}^g)^2/(2\theta^g)]$, $(h')^{eq}=h^{eq}(\vec{r},\vec{v}',t)=\rho^h(2\pi \theta^h)^{-3/2}\exp[-(\vec{v}'-\vec{u}^h)^2/(2\theta^h)]$. Notice that the stationary solutions for $(g')^{eq}$ and $(h')^{eq}$ both adopt a gaussian form with respect to the $\vec{v}'$ up to the second order. 

So that under the BGK approximation, our above derivation for the classical Boltzmann equation still applies to the Hamilton-Boltzmann equations, which leads to
\begin{eqnarray}
	\frac{Dg}{Dt}&=&\int\int d\vec{v}_1d\vec{\omega}B(\vec{\omega}) [h_1^{eq}g^{eq}g_1^{eq}-h_1gg_1]\nonumber\\
&=&g^{eq}\int d\vec{v}_1  h_1^{eq}g_1^{eq}\cdot\int d\vec{\omega}B(\vec{\omega})-g\int d\vec{v}_1  h_1g_1\cdot\int d\vec{\omega}B(\vec{\omega}) =\frac{\rho^f}{\sigma}(g^{eq}-g),\\
	\frac{Dh}{Dt}&=&-\int\int d\vec{v}_1d\vec{\omega}B(\vec{\omega})[g_1^{eq}h^{eq}h_1^{eq}-g_1hh_1]\nonumber\\
&=&-h^{eq}\int d\vec{v}_1  h_1^{eq}g_1^{eq}\cdot\int d\vec{\omega}B(\vec{\omega})+h\int d\vec{v}_1  h_1g_1\cdot\int d\vec{\omega}B(\vec{\omega})=-\frac{\rho^f}{\sigma}(h^{eq}-h),
\end{eqnarray}
where $\sigma^{-1}=\int d{\vec{\omega}}B(\vec{\omega})$ is the collision frequency. $\rho^f(\vec{r},t)=\int h_1g_1d\vec{v}_1=\int (h_1)^{eq}(g_1)^{eq}d\vec{v}_1$ denotes the particle number density.

We call the above equations as the HB-BGK model. And it is straightforward to check that the HB-BGK model is still a Hamiltonian system, whose Hamiltonian function is given by
\begin{eqnarray}
H(g,h)=-\int\int d\vec{r}d\vec{v}\bigg(\vec{v}\cdot\frac{\partial g}{\partial \vec{r}}\bigg)h+\frac{1}{\sigma}\int\int d\vec{r}d\vec{v}\rho^f\big(hg^{eq}+h^{eq}g-hg-h^{eq}g^{eq}\big).
\end{eqnarray}
The above formula can be derived from the original one $H(g,h)=-\int\int d\vec{r}d\vec{v}(\vec{v}\cdot\partial g/\partial \vec{r})h-\int\int\int d\vec{r}d\vec{v}d\vec{v_1}\big(hh_1-h^{eq}h_1^{eq}\big)\big(gg_1-g^{eq}g_1^{eq}\big)/(4\sigma)$. 

With respect to variables $f=gh$ and $y=\ln(h)$, the BH-BGK model can be transformed into an alternative equilvalent form, i.e.
\begin{eqnarray}
	\frac{Df}{Dt}&=&\frac{\rho^f(\vec{r},t)}{\sigma}\big[\exp(y-y^{eq})f^{eq}-\exp(y^{eq}-y)f\big],\\
	\frac{Dy}{Dt}&=&-\frac{\rho^f(\vec{r},t)}{\sigma}[\exp(y^{eq}-y)-1].
\end{eqnarray}
Thanks to the nature of canonical transformation, it is still a Hamiltonian system with the Hamiltonian function as $H(f,y)=-\int\int d\vec{r}d\vec{v}\big(\vec{v}\cdot\frac{\partial f}{\partial \vec{r}}\big)y+\frac{1}{\sigma}\int\int d\vec{r}d\vec{v}\rho^f\big[\exp(y-y^{eq})f^{eq}+\exp(y^{eq}-y)f-f-f^{eq}\big]$.


\subsection{Hydrodynamic Limits for the HB-BGK model}
It is well-known that the Boltzmann equation is a mesoscopic model, based on which hydrodynamic models in the macroscopic scale can be rigorously derived. Classical methods include the Chapman-Enskog expansion, Grad's moment closure and Levermore's moment closure, etc. Here we apply the Grad's moment closure method to the HB-BGK model and expect more general hydrodynamic models beyond the Euler or Navier-Stokes equations can be obtained.

Focus on the three dimensional case. We multiply $\chi=1, v_i, v_iv_j, v_iv_kv_k$ on both sides of the HB-BGK model and adopt the Einstein summation convection. Then a coupled model for 13 moments can be derived (see Appendix for details), i.e.
\begin{eqnarray*}
&&\frac{\partial \rho^g}{\partial t}+\frac{\partial(\rho^g u_k^g)}{\partial x_k}=0,\\
&&\frac{\partial (\rho^g u_i^g)}{\partial t}+\frac{\partial(\rho^g u_i^gu_k^g+\rho^g\theta_{ik}^g)}{\partial x_k}=0,\\
&&\frac{\partial}{\partial t}\theta_{ij}^g+u_k^g\frac{\partial}{\partial x_k}\theta_{ij}^g+\theta_{jk}^g\frac{\partial}{\partial x_k}u_i^g+\theta_{ik}^g\frac{\partial}{\partial x_k}u_j^g+\frac{2}{5\rho^g}\bigg[\frac{\partial}{\partial x_i}q_j^g+\frac{\partial}{\partial x_j}q_i^g+\delta_{ij}\frac{\partial}{\partial x_k}q_k^g\bigg]\\
&&=-\frac{\rho^f(\theta_{ij}^g-\theta^g\delta_{ij})}{\sigma},\\
&&\frac{\partial}{\partial t}q_i^g+u_k^g\frac{\partial}{\partial x_k}q_i^g-[\theta_{ij}^g\theta_{jk}^g-2\theta^g\theta_{ik}^g+(\theta^g)^2\delta_{ik}]\frac{\partial}{\partial x_k}\rho^g+\frac{7}{5}q_i^g\frac{\partial}{\partial x_k}u_k^g+\frac{7}{5}q_k^g\frac{\partial}{\partial x_k}u_i^g+\frac{2}{5}q_k^g\frac{\partial}{\partial x_i}u_k^g,\\
&&-\rho^f\theta_{ik}^g\bigg(\frac{\partial}{\partial x_j}\theta_{jk}^g-\frac{7}{2}\frac{\partial}{\partial x_k}\theta^g\bigg)+2\rho^g\theta^g\bigg(\frac{\partial}{\partial x_k}\theta_{ik}^g-\frac{\partial}{\partial x_i}\theta^g\bigg)=-\frac{\rho^f[q_i^g+\rho^gu_j^g(\theta_{ij}^g-\theta^g\delta_{ij})]}{\sigma},
\end{eqnarray*}
\begin{eqnarray*}
&&\frac{\partial \rho^h}{\partial t}+\frac{\partial(\rho^h u_k^h)}{\partial x_k}=0,\\
&&\frac{\partial (\rho^h u_i^h)}{\partial t}+\frac{\partial(\rho^h u_i^hu_k^h+\rho^h\theta_{ik}^h)}{\partial x_k}=0,\\
&&\frac{\partial}{\partial t}\theta_{ij}^h+u_k^h\frac{\partial}{\partial x_k}\theta_{ij}^h+\theta_{jk}^h\frac{\partial}{\partial x_k}u_i^h+\theta_{ik}^h\frac{\partial}{\partial x_k}u_j^h+\frac{2}{5\rho^h}\bigg[\frac{\partial}{\partial x_i}q_j^h+\frac{\partial}{\partial x_j}q_i^h+\delta_{ij}\frac{\partial}{\partial x_k}q_k^h\bigg]=\frac{\rho^f(\theta_{ij}^h-\theta^h\delta_{ij})}{\sigma},\\
&&\frac{\partial}{\partial t}q_i^h+u_k^h\frac{\partial}{\partial x_k}q_i^h-[\theta_{ij}^h\theta_{jk}^h-2\theta^h\theta_{ik}^h+(\theta^h)^2\delta_{ik}]\frac{\partial}{\partial x_k}\rho^h+\frac{7}{5}q_i^h\frac{\partial}{\partial x_k}u_k^h+\frac{7}{5}q_k^h\frac{\partial}{\partial x_k}u_i^h+\frac{2}{5}q_k^h\frac{\partial}{\partial x_i}u_k^h\\
&&-\rho^h\theta_{ik}^h\bigg(\frac{\partial}{\partial x_j}\theta_{jk}^h-\frac{7}{2}\frac{\partial}{\partial x_k}\theta^h\bigg)+2\rho^h\theta^h\bigg(\frac{\partial}{\partial x_k}\theta_{ik}^h-\frac{\partial}{\partial x_i}\theta^h\bigg)=\frac{\rho^f[q_i^h+\rho^h u_j^h(\theta_{ij}^h-\theta^h\delta_{ij})]}{\sigma},
\end{eqnarray*}
where $\rho^g(\vec{r},t)=\int g d\vec{v}$, $\rho^gu_i^g(\vec{r},t)=\int gv_i d\vec{v}$, $\rho^g \theta^g(\vec{r},t)=\frac{1}{3}\int g(v_k-u_k^g)(v_k-u_k^g) d\vec{v}$, $\rho^g\theta_{ij}^g=\int g(v_i-u_i^g)(v_j-u_j^g) d\vec{v}$, $q_i^g=\frac{1}{2}\int g(v_i-u_i^g)(v_k-u_k^g)(v_k-u_k^g) d\vec{v}$. Similar variables are defined for the velocity integral with respect to function $h$, i.e. $\rho^h(\vec{r},t)=\int h d\vec{v}$, $\rho^hu_i^h(\vec{r},t)=\int hv_i d\vec{v}$, $\rho^h \theta^h(\vec{r},t)=\frac{1}{3}\int h(v_k-u_k^h)(v_k-u_k^h) d\vec{v}$, $\rho^h\theta_{ij}^h=\int h(v_i-u_i^h)(v_j-u_j^h) d\vec{v}$, $q_i^h=\frac{1}{2}\int h(v_i-u_i^h)(v_k-u_k^h)(v_k-u_k^h) d\vec{v}$. Notice that the above equations are weakly coupled through the factor $\rho^f$, which is expressed through 
\begin{eqnarray}
\rho^f=\frac{\rho^g\rho^h}{(2\pi)^{3/2}(\theta^g+\theta^h)^{3/2}}\exp\bigg[-\frac{(u_k^g-u_k^h)(u_k^g-u_k^h)}{2(\theta^g+\theta^h)}\bigg].
\end{eqnarray}

By replacing $g$ and $h$ with their respect equilibrium solutions $g^{eq}$ and $h^{eq}$, terms on the right-hand sides of above continuum equations all become zeros, so does the heat flux itself. This means the first three equations are decoupled from the last three ones and both of them become the classical Euler equations. For this reason, we will drop the superscripts $g$ and $h$ for simplicity.
\begin{eqnarray*}
&&\frac{\partial \rho}{\partial t}+\nabla\cdot(\rho\vec{u})=0,\\
&&\frac{\partial\vec{u}}{\partial t}+\vec{u}\cdot\nabla\vec{u}=-\frac{\nabla(\rho \theta)}{\rho},\\
&&\frac{\partial}{\partial t}\theta+\vec{u}\cdot\nabla \theta=-\frac{2}{3}\theta\nabla\cdot\vec{u}.
\end{eqnarray*}
It is easy to check that the entropy function $S(\rho,\theta)=-3\int d\vec{r}\rho[\ln\theta-(2\ln\rho)/3+\ln(2\pi)+1]/2$ is time-conserved for above Euler equations.

\section{Conclusion and Discussion}

The Boltzmann equation plays a significant role in many fields, including gas dynamics, plasma kinetics, heat conduction, etc. However, due to its relatively complicated collision kernel, many basic mathematical properties of the Boltzmann equation have not been justified yet.

In the current study, by introducing a stochastic picture of the binary collision of particles, we reformulate the Boltzmann equation into a broader Markov-chain model. The corresponding Kolmogorov forward equations and Liouville equation in either discrete or continuous time and state space are constructed respectively, both of which offer stochastic generalizations of the classical Boltzmann equation. Then by using the WKB method, the large deviations principle for the fluctuating Boltzmann equation is constructed, which on one hand explains the probabilistic origin of the H-theorem, on the other hand provides time-reversible generalizations of the Boltzmann equation in a Hamiltonian structure. In a similar way, the diffusive limit is discussed too.

During the preparation of this manuscript, we notice a closely related work published by Prof. F. Bouchet\cite{Bouchet-2020}, which considered the kinetic theory of dilute gases in the Boltzmann–Grad limit and made a large deviation estimate for the probability of the empirical distribution. The large deviations rate function he obtained coincides with ours. But he focused on the gradient structure for the Boltzmann equation, in contrast to the Hamiltonian structure we considered. Furthermore, the Liouville equation, the Hamilton-Jacobi equation as well as results on the diffusive limit were not included in his paper.

\section*{Declaration of Competing Interest}
The authors declare that they have no confict of interest.

\section*{Funding}
This work was supported by the National Key R\&D Program of China
(2023YFC2308702) and Guangdong Basic and Applied Basic Research Foundation (2023A1515010157). The author thanks the hospitality of Beijing Institute of Mathematical Sciences and Applications (BIMSA), and thank Prof. Hong Qian, Wieran Sun, Jingwei Hu and Liangrong Peng for their stimulating discussions.

\section*{Appendix: Grad's 13 moments equations}
Here we firstly represent the results of Grad for the BGK model. The derivation for the HB-BGK model is similar.  We start with the equilibrium solution of the Boltzmann equation, which reads $f^{eq}(\vec{r},\vec{v},t)=\rho(2\pi \theta)^{-3/2}\exp[-(\vec{v}-\vec{u})^2/(2\theta)]$, where  $\rho(\vec{r},t)=\int f d\vec{v}$, $\rho\vec{u}(\vec{r},t)=\int\vec{v}f d\vec{v}$ and $3\rho \theta(\vec{r},t)/2=\int (\vec{v}-\vec{u})^2f d\vec{v}/2$ represent the local particle number density, momentum and kinetic energy in the three dimensional space separately.

Approximate the distribution function around the equilibrium state by Hermite expansion, i.e. $f(\vec{r},\vec{v},t)=f^{eq}\sum_{n=1}^N\sum_{i_1,i_2\cdots,i_n}\alpha^{(n)}_{i_1,i_2\cdots,i_n}(\vec{r},t) He_{i_1,i_2\cdots,i_n}^{(n)}(\vec{C})$, where $\vec{C}=\theta^{-1/2}(\vec{v}-\vec{u})$, $i_1.\cdots,i_n\in\{x,y,z\}$. $\alpha^{(n)}_{i_1,i_2\cdots,i_n}$ are tensorial coefficients depending on $\vec{r}$ and $t$. The $n$'th order tensorial Hermite polynomials are defined through
\begin{eqnarray*}
He_{i_1,i_2\cdots,i_n}^{(n)}(\vec{C})=(-1)^n\frac{\partial^n(\ln f^{eq})}{\partial C_{i_1}\cdots \partial C_{i_n}},
\end{eqnarray*}
which satisfy the orthogonal condition with respect to each other, i.e.
\begin{eqnarray*}
\int d\vec{C}He^{(m)}_{i_1,i_2\cdots,i_m}(\vec{C})He^{(n)}_{j_1,j_2\cdots,j_n}(\vec{C})f^{eq}(\vec{r},\vec{v},t)=\rho\theta^{-3/2}\delta_{mn}\sum_{j_1',\cdots,j_n'}\delta_{i_1j_1'}\cdots\delta_{i_nj_n'}. \end{eqnarray*}
$j_1',\cdots,j_n'$ stands for all possible permutations of $j_1,\cdots,j_n$.
Without further mention, the Einstein summation convention for repeated subscript is adopted automatically.

Inserting the approximate distribution function into the BGK model, the Grad's moment closure equations for hydrodynamics are obtained. In particular, expand the polynomial space up to the third order as $\{1,v_i,v_iv_j,v_iv_kv_k\}$, and introduce the stress tensor $\rho\theta_{ij}=\theta\int C_iC_j f d\vec{v}$ (in particular $\theta=\theta_{ii}/3$) and heat flux $q_i=\frac{1}{2}\theta^{3/2}\int C_iC_kC_kfd\vec{v}$. We arrive at Grad's 13 moments equations:
\begin{eqnarray*}
&&\frac{\partial \rho}{\partial t}+\frac{\partial(\rho u_k)}{\partial x_k}=\frac{D\rho}{Dt}+\rho\frac{\partial u_k}{\partial x_k}=0,\\
&&\frac{\partial (\rho u_i)}{\partial t}+\frac{\partial(\rho u_iu_k+\rho\theta_{ik})}{\partial x_k}=\frac{D u_i}{Dt}+\theta_{ik}\frac{\partial \ln\rho}{\partial x_k}+\frac{\partial \theta_{ik}}{\partial x_k}=0,\\
&&\frac{D}{Dt}\theta_{ij}+\theta_{jk}\frac{\partial}{\partial x_k}u_i+\theta_{ik}\frac{\partial}{\partial x_k}u_j+\frac{2}{5\rho}\bigg[\frac{\partial}{\partial x_i}q_j+\frac{\partial}{\partial x_j}q_i+\delta_{ij}\frac{\partial}{\partial x_k}q_k\bigg]=-\frac{\rho(\theta_{ij}-\theta\delta_{ij})}{\sigma}\\
&&\frac{D}{Dt}q_i-(\theta_{ij}\theta_{jk}-2\theta\theta_{ik}+\theta^2\delta_{ik})\frac{\partial}{\partial x_k}\rho+\frac{7}{5}q_i\frac{\partial}{\partial x_k}u_k+\frac{7}{5}q_k\frac{\partial}{\partial x_k}u_i+\frac{2}{5}q_k\frac{\partial}{\partial x_i}u_k\\
&&-\rho\theta_{ik}\bigg(\frac{\partial}{\partial x_j}\theta_{jk}-\frac{7}{2}\frac{\partial}{\partial x_k}\theta\bigg)+2\rho\theta\bigg(\frac{\partial}{\partial x_k}\theta_{ik}-\frac{\partial}{\partial x_i}\theta\bigg)=-\frac{\rho[q_i+\rho u_j(\theta_{ij}-\theta\delta_{ij})]}{\sigma}
\end{eqnarray*}
where $D/Dt=\partial/\partial t+u_k\partial/\partial x_k$. The details of derivation can be found in what follows.

\textbf{(1) Construction of the non-equilibrium distribution function.}

With respect to the 13 Hermite polynomials in the space of $\{1,v_i,v_iv_j,v_iv_kv_k\}$, the non-equilibrium distribution function reads
\begin{eqnarray*}
f(t,\vec{r},\vec{v})=f^{eq}\big[\alpha^{(0)}He^{(0)}(\vec{C})+\alpha^{(1)}_iHe^{(1)}_i(\vec{C})+\alpha^{(2)}_{ij}He^{(2)}_{ij}(\vec{C})+\alpha^{(3)}_{ikk}He^{(3)}_{ill}(\vec{C})\big]\\
\end{eqnarray*}
where $He^{(0)}(\vec{C})=1$, $He^{(1)}_i(\vec{C})=C_i$, $He^{(2)}_{ij}(\vec{C})=C_iC_j-\delta_{ij}$, $He^{(3)}_{ijk}(\vec{C})=C_iC_jC_k-C_i\delta_{jk}-C_j\delta_{ik}-C_k\delta_{ij}$. Further making use of the definitions of $\rho$, $u_i$, $\theta_{ij}$ and $q_i$, it becomes possible to derive $\alpha^{(0)}=1$, $\alpha^{(1)}_i=0$, $\alpha^{(2)}_{ij}=(\theta_{ij}-\delta_{ij}\theta)/(2\theta)$, $\alpha^{(3)}_{ikk}=q_i/(5\rho\theta^{3/2})$. It means
\begin{eqnarray*}
f(t,\vec{r},\vec{v})=f^{eq}\bigg[1+\frac{1}{2\theta}\big(\theta_{ij}-\delta_{ij}\theta\big)C_iC_j+\frac{q_i}{5\rho\theta^{3/2}}C_i\big(C_kC_k-5\big)\bigg].
\end{eqnarray*}

\textbf{(2) Calculating the time evolution of stress tensors.}

Define $\vec{\xi}=\vec{v}-\vec{u}$. Then we have
\begin{eqnarray*}
&&\int d\vec{v} v_iv_j\frac{\partial}{\partial t}f+\int d\vec{v} v_iv_jv_k\frac{\partial}{\partial x_k}f=\frac{\partial}{\partial t}\int d\vec{v} v_iv_jf+\frac{\partial}{\partial x_k}\int d\vec{v} v_iv_jv_kf\\
&=&\frac{\partial}{\partial t}\int d\vec{v} (\xi_i\xi_j+u_i\xi_j+u_j\xi_i+u_iu_j)f\\
&&+\frac{\partial}{\partial x_k}\int d\vec{v} (\xi_i\xi_j\xi_k+u_i\xi_j\xi_k+u_j\xi_i\xi_k+u_iu_j\xi_k+\xi_i\xi_ju_k+u_i\xi_ju_k+u_j\xi_iu_k+u_iu_ju_k)f\\
&=&\frac{\partial}{\partial t}(\rho\theta_{ij}+\rho u_iu_j)+\frac{\partial}{\partial x_k}(q_{ijk}+\rho u_i\theta_{jk}+\rho u_j\theta_{ik}+\rho u_k\theta_{ij}+\rho u_iu_ju_k)\\
&=&\rho\frac{\partial}{\partial t}\theta_{ij}+\theta_{ij}\frac{\partial}{\partial t}\rho+u_j\frac{\partial}{\partial t}(\rho u_i)+\rho u_i\frac{\partial}{\partial t}u_j+\rho u_i\frac{\partial}{\partial x_k}\theta_{jk}+\theta_{jk}\frac{\partial}{\partial x_k}(\rho u_i)+\rho u_j\frac{\partial}{\partial x_k}\theta_{ik}\\
&&+\theta_{ik}\frac{\partial}{\partial x_k}(\rho u_j)+\rho u_k\frac{\partial}{\partial x_k}\theta_{ij}+\theta_{ij}\frac{\partial}{\partial x_k}(\rho u_k)+\rho u_iu_k\frac{\partial}{\partial x_k}u_j+u_j\frac{\partial}{\partial x_k}(\rho u_iu_k)+\frac{\partial}{\partial x_k}q_{ijk}\\
&=&u_j\bigg[\frac{\partial}{\partial t}(\rho u_i)+\frac{\partial}{\partial x_k}(\rho\theta_{ik})+\frac{\partial}{\partial x_k}(\rho u_iu_k)\bigg]+\rho u_i\bigg[\frac{\partial}{\partial t}u_j+u_k\frac{\partial}{\partial x_k}u_j+\theta_{jk}\frac{\partial}{\partial x_k}\ln\rho+\frac{\partial}{\partial x_k}\theta_{jk}\bigg]\\
&&+\theta_{ij}\bigg[\frac{\partial}{\partial t}\rho+\frac{\partial}{\partial x_k}(\rho u_k)\bigg]+\rho\frac{\partial}{\partial t}\theta_{ij}+\rho u_k\frac{\partial}{\partial x_k}\theta_{ij}+\rho\theta_{jk}\frac{\partial}{\partial x_k}u_i+\rho\theta_{ik}\frac{\partial}{\partial x_k}u_j+\frac{\partial}{\partial x_k}q_{ijk}\\
&=&\rho\frac{\partial}{\partial t}\theta_{ij}+\rho u_k\frac{\partial}{\partial x_k}\theta_{ij}+\rho\theta_{jk}\frac{\partial}{\partial x_k}u_i+\rho\theta_{ik}\frac{\partial}{\partial x_k}u_j+\frac{\partial}{\partial x_k}q_{ijk},
\end{eqnarray*}
by noticing terms in above three brackets all exactly cancelled by applying the balance equations for the number density and momentum. 

\textbf{(3) Calculating the time evolution of heat fluxes.}
\begin{eqnarray*}
&&\int d\vec{v} v_iv_jv_j\frac{\partial}{\partial t}f+\int d\vec{v} v_iv_jv_jv_k\frac{\partial}{\partial x_k}f=\frac{\partial}{\partial t}\int d\vec{v} v_iv_jv_jf+\frac{\partial}{\partial x_k}\int d\vec{v} v_iv_jv_jv_kf\\
&=&\frac{\partial}{\partial t}\int d\vec{v} (\xi_j\xi_j\xi_i+2u_j\xi_j\xi_i+u_ju_j\xi_i+\xi_j\xi_ju_i+2u_j\xi_ju_i+u_ju_ju_i)f\\
&&+\frac{\partial}{\partial x_k}\int d\vec{v} (\xi_j\xi_j\xi_i\xi_k+\xi_j\xi_ju_i\xi_k+\xi_j\xi_j\xi_iu_k+\xi_j\xi_ju_iu_k+2u_j\xi_j\xi_i\xi_k+u_ju_j\xi_i\xi_k\\
&&+2u_j\xi_ju_i\xi_k+u_ju_ju_i\xi_k+2u_j\xi_j\xi_iu_k+u_ju_j\xi_iu_k+2u_j\xi_ju_iu_k+u_ju_ju_iu_k)f\\
&=&\frac{\partial}{\partial t}(2q_i+2\rho u_j\theta_{ij}+\rho u_i\theta_{jj}+\rho u_iu_ju_j)+\frac{\partial}{\partial x_k}(\omega_{ik}+2u_iq_k+2u_kq_i+\rho u_iu_k\theta_{jj}+2u_jq_{ijk}\\
&&+\rho u_iu_j\theta_{ik}+2\rho u_iu_j\theta_{jk}+2\rho u_ju_k\theta_{ij}+\rho u_iu_ku_ju_j)\\
&=&2\frac{\partial}{\partial t}q_i+2\rho u_j\frac{\partial}{\partial t}\theta_{ij}+2\theta_{ij}\frac{\partial}{\partial t}(\rho u_j)+\rho u_i\frac{\partial}{\partial t}\theta_{jj}+\theta_{jj}\frac{\partial}{\partial t}(\rho u_i)+u_ju_j\frac{\partial}{\partial t}(\rho u_i)+2\rho u_iu_j\frac{\partial}{\partial t}u_j\\
&&+\frac{\partial}{\partial x_k}w_{ik}+2u_i\frac{\partial}{\partial x_k}q_k+2q_k\frac{\partial}{\partial x_k}u_i+2u_k\frac{\partial}{\partial x_k}q_i+2q_i\frac{\partial}{\partial x_k}u_k+\rho u_iu_k\frac{\partial}{\partial x_k}\theta_{jj}+\rho u_i\theta_{jj}\frac{\partial}{\partial x_k}u_k\\
&&+u_k\theta_{jj}\frac{\partial}{\partial x_k}(\rho u_i)+2u_j\frac{\partial}{\partial x_k}q_{ijk}+2q_{ijk}\frac{\partial}{\partial x_k}u_j +u_ju_j\frac{\partial}{\partial x_k}(\rho\theta_{ik})+ 2\rho u_j\theta_{ik}\frac{\partial}{\partial x_k}u_j\\
&&+2\rho u_iu_j\frac{\partial}{\partial x_k}\theta_{jk}+2\rho u_i\theta_{jk}\frac{\partial}{\partial x_k}u_j+2 u_j\theta_{jk}\frac{\partial}{\partial x_k}(\rho u_i)+2\rho u_ju_k\frac{\partial}{\partial x_k}\theta_{ij}+2\rho u_j\theta_{ij}\frac{\partial}{\partial x_k}u_k\\
&&+2u_k\theta_{ij}\frac{\partial}{\partial x_k}(\rho u_j)+u_ku_ju_j\frac{\partial}{\partial x_k}(\rho u_i)+2\rho u_iu_ju_k\frac{\partial}{\partial x_k}u_j+\rho u_iu_ju_j\frac{\partial}{\partial x_k}u_k\\
&=&\rho u_i\bigg[\frac{\partial}{\partial t}\theta_{jj}+u_k\frac{\partial}{\partial x_k}\theta_{jj}+2\theta_{jk}\frac{\partial}{\partial x_k}u_j+\frac{2}{\rho}\frac{\partial}{\partial x_k}q_k\bigg]+\theta_{jj}\bigg[\frac{\partial}{\partial t}(\rho u_i)+\frac{\partial}{\partial x_k}(\rho u_iu_k)+\frac{\partial}{\partial x_k}(\rho\theta_{ik})\bigg]\\
&&+u_ju_j\bigg[\frac{\partial}{\partial t}(\rho u_i)+\frac{\partial}{\partial x_k}(\rho u_i u_k)+\frac{\partial}{\partial x_k}(\rho\theta_{ik})\bigg]+2\theta_{ij}\bigg[\frac{\partial}{\partial t}(\rho u_j)+\frac{\partial}{\partial x_k}(\rho u_ju_k)+\frac{\partial}{\partial x_k}(\rho\theta_{jk})\bigg]\\
&&+2\rho u_iu_j\bigg[\frac{\partial}{\partial t}u_j+u_k\frac{\partial}{\partial x_k}u_j+\theta_{jk}\frac{\partial}{\partial x_k}\ln\rho+\frac{\partial}{\partial x_k}\theta_{jk}\bigg]+2\rho u_j\bigg[\frac{\partial}{\partial t}\theta_{ij}+u_k\frac{\partial}{\partial x_k}\theta_{ij}+\theta_{ik}\frac{\partial}{\partial x_k}u_j\\
&&+\theta_{jk}\frac{\partial}{\partial x_k}u_i+\rho^{-1}\frac{\partial}{\partial x_k}q_{ijk}\bigg]+2\frac{\partial}{\partial t}q_i+2u_k\frac{\partial}{\partial x_k}q_i+2q_k\frac{\partial}{\partial x_k}u_i+2q_i\frac{\partial}{\partial x_k}u_k+\frac{\partial}{\partial x_k}w_{ik}\\
&&+2q_{ijk}\frac{\partial}{\partial x_k}u_j-\theta_{jj}\frac{\partial}{\partial x_k}(\rho\theta_{ik})-2\theta_{ij}\frac{\partial}{\partial x_k}(\rho\theta_{jk})\\
&=&2\rho u_j\bigg[\frac{\partial}{\partial t}\theta_{ij}+u_k\frac{\partial}{\partial x_k}\theta_{ij}+\theta_{ik}\frac{\partial}{\partial x_k}u_j+\theta_{jk}\frac{\partial}{\partial x_k}u_i+\rho^{-1}\frac{\partial}{\partial x_k}q_{ijk}\bigg]+2\frac{\partial}{\partial t}q_i+2u_k\frac{\partial}{\partial x_k}q_i\\
&&+2q_k\frac{\partial}{\partial x_k}u_i+2q_i\frac{\partial}{\partial x_k}u_k+\frac{\partial}{\partial x_k}w_{ik}+2q_{ijk}\frac{\partial}{\partial x_k}u_j-\theta_{jj}\frac{\partial}{\partial x_k}(\rho\theta_{ik})-2\theta_{ij}\frac{\partial}{\partial x_k}(\rho\theta_{jk}).\\
\end{eqnarray*}
Again, we apply the balance equations for the number density, momentum and kinetic energy in the last step.

\textbf{(4) Calculation of the third-order and fourth-order moments.}

\begin{eqnarray*}
q_{ijk}&=&\int \xi_i\xi_j\xi_kfd\vec{v}=\theta^{3}\int [He^{(3)}_{ijk}(\vec{C})+He^{(1)}_i(\vec{C})\delta_{jk}+He^{(1)}_j(\vec{C})\delta_{ik}+He^{(1)}_k(\vec{C})\delta_{ij}]fd\vec{C}\\
&=&\theta^{3}\alpha^{(3)}_{lmm}\int He^{(3)}_{ijk}(\vec{C})He^{(3)}_{lnn}(\vec{C})f^{eq}d\vec{C}=\frac{2}{5}q_l(\delta_{il}\delta_{jn}\delta_{kn}+\delta_{in}\delta_{jl}\delta_{kn}+\delta_{in}\delta_{jn}\delta_{kl})\\
&=&\frac{2}{5}(q_i\delta_{jk}+q_j\delta_{ik}+q_k\delta_{ij}),\\
w_{ij}&=&\int \xi_i\xi_j\xi_k\xi_kfd\vec{v}\\
&=&\theta^{7/2}\int \big[He^{(4)}_{ijkk}(\vec{C})+He^{(2)}_{ij}(\vec{C})\delta_{kk}+2He^{(2)}_{ik}(\vec{C})\delta_{jk}+2He^{(2)}_{jk}(\vec{C})\delta_{ik}
+He^{(2)}_{kk}(\vec{C})\delta_{ij}+5\delta_{ij}\big]fd\vec{C}\\
&=&\theta^{7/2}\alpha^{(2)}_{mn}\int \big[He^{(2)}_{ij}(\vec{C})\delta_{kk}+2He^{(2)}_{ik}(\vec{C})\delta_{jk}+2He^{(2)}_{jk}(\vec{C})\delta_{ik}+He^{(2)}_{kk}(\vec{C})\delta_{ij}\big]He^{(2)}_{mn}(\vec{C})f^{eq}d\vec{C}\\
&&+5\rho \theta^2\delta_{ij}=\rho\theta^{2}(14\alpha^{(2)}_{ij}+2\alpha^{(2)}_{kk}\delta_{ij})+5\rho \theta^2\delta_{ij}=\rho\theta(7\theta_{ij}-2\theta\delta_{ij}),
\end{eqnarray*}
by using the orthogonal conditions of Hermite polynomials as well as the fourth order Hermite polynomials $He^{(4)}_{ijkl}(\vec{C})=C_iC_jC_kC_l-(C_iC_j\delta_{kl}+C_iC_k\delta_{jl}+C_iC_l\delta_{jk}+C_jC_k\delta_{il}+C_jC_l\delta_{ik}+C_kC_l\delta_{ij})+(\delta_{ij}\delta_{kl}+\delta_{ik}\delta_{jl}+\delta_{il}\delta_{jk})$.


\textbf{(5) Calculating the right-hand side terms of the BGK model.}

Direct calculation of the moment integral of the collision kernel in the original Boltzmann equation is cumbersome. Here we turn to a simple case -- its linear approximation in the BGK model. It is straightforward to verify that $1,\vec{v},\vec{v}^2$ are three collision invariants, which means their velocity integrals with respect to the collision kernel in the BGK model are all zeros.
\begin{eqnarray*}
&&\int d\vec{v}(f^{eq}-f)=\rho-\rho=0,\\
&&\int d\vec{v}(f^{eq}-f)v_i=\rho u_i-\rho u_i=0,\\
&&\int d\vec{v}(f^{eq}-f)v_iv_i=(3\rho\theta+\rho u_iu_i)-(3\rho\theta+\rho u_iu_i)=0.
\end{eqnarray*}
So that we only need to show
\begin{eqnarray*}
&&\int d\vec{v}(f^{eq}-f)v_iv_j=\int d\vec{v}(f^{eq}-f)(\xi_i+u_i)(\xi_j+u_j)=-\rho(\theta_{ij}-\theta\delta_{ij}),\\
&&\int d\vec{v}(f^{eq}-f)v_iv_jv_j=\int d\vec{v}(f^{eq}-f)(\xi_i+u_i)(\xi_j+u_j)(\xi_j+u_j)\\
&=&\int d\vec{v}(f^{eq}-f)(\xi_i\xi_j\xi_j+u_i\xi_j\xi_j+2u_j\xi_i\xi_j+\xi_iu_ju_j+2\xi_ju_iu_j+u_iu_ju_j)\\
&=&-2q_i-2\rho u_j(\theta_{ij}-\theta\delta_{ij}).
\end{eqnarray*}

\textbf{(6) Expressing the density function $\rho^f$.}

In the BH-BGK model, two equations are coupled through a common factor $\rho^f$, which is the time integral of the equilibrium distribution $f^{eq}=h^{eq}g^{eq}$. Due to the well-known fact that the product of two gaussion distributions is still a gaussian distribution, we have
\begin{eqnarray*}
h^{eq}g^{eq}&=&\frac{\rho^g}{(2\pi\theta^g)^{3/2}}\exp\bigg[-\frac{(v_k-u_k^g)(v_k-u_k^g)}{2\theta^g}\bigg]\cdot \frac{\rho^h}{(2\pi\theta^h)^{3/2}}\exp\bigg[-\frac{(v_k-u_k^h)(v_k-u_k^h)}{2\theta^h}\bigg]\\
&=&\frac{\rho^f}{(2\pi\theta^f)^{3/2}}\exp\bigg[-\frac{(v_k-u_k^f)(v_k-u_k^f)}{2\theta^f}\bigg]=f^{eq},
\end{eqnarray*} 
where the particle number density, velocity and effective temperature read
\begin{eqnarray*}
\rho^f&=&\frac{\rho^g\rho^h}{(2\pi)^{3/2}(\theta^g+\theta^h)^{3/2}}\exp\bigg[-\frac{(u_k^g-u_k^h)(u_k^g-u_k^h)}{2(\theta^g+\theta^h)}\bigg],\\
u_k^f&=&\frac{u_k^g\theta^h+u_k^h\theta^g}{\theta^g+\theta^h},\\
\theta^f&=&\frac{\theta^g\theta^h}{\theta^g+\theta^h}.
\end{eqnarray*}
Above formulas allow us to express the unknown factor $\rho^f$ through known variables with respect to functions $g$ and $h$.

\end{document}